\documentclass[a4paper,11pt]{article}
\usepackage{a4wide}
\usepackage{amssymb}
\usepackage{amsmath}
\usepackage[english]{babel}
\usepackage{sidecap}
\usepackage{subfigure}
\usepackage{here}
\usepackage{graphicx}
\usepackage{color}
\usepackage{listings}
\usepackage{epstopdf}
\usepackage{authblk}
\usepackage[numbers]{natbib}

\usepackage[pdftitle={Mesoscopic Modeling of Random Walk and Reactions in
  Crowded Media},
  pdfauthor={Engblom/Lotstedt/Meinecke},
  pdffitwindow=true,
  breaklinks=true,
  colorlinks=true,
  urlcolor=blue,
  linkcolor=red,
  citecolor=red,
  anchorcolor=red]{hyperref}

\numberwithin{equation}{section}
\numberwithin{table}{section}
\numberwithin{figure}{section}

\def\calI{\mathcal{I}}
\def\calJ{\mathcal{J}}

\def\calM{\mathcal{M}}
\def\calN{\mathcal{N}}
\def\calO{\mathcal{O}}

\def\calT{\mathcal{T}}
\def\calV{\mathcal{V}}
\def\calW{\mathcal{W}}
\def\fatA{\mathbf{A}}
\def\fatB{\mathbf{B}}

\def\fatG{\mathbf{G}}
\def\fatH{\mathbf{H}}
\def\fatD{\mathbf{D}}

\def\fate{\mathbf{e}}
\def\fatp{\mathbf{p}}
\def\fatf{\mathbf{f}}
\def\fatg{\mathbf{g}}

\def\fatI{\mathbf{I}}

\def\fatM{\mathbf{M}}
\def\fatn{\mathbf{n}}

\def\fatu{\mathbf{u}}

\def\fatx{\mathbf{x}}

\def\fatq{\mathbf{q}}

\def\fatR{\mathbf{R}}

\def\fatS{\mathbf{S}}

\def\fatT{\mathbf{T}}
\def\fatU{\mathbf{U}}

\def\fatv{\mathbf{v}}

\def\fatx{\mathbf{x}}
\def\faty{\mathbf{y}}
\def\fatby{\bar{\mathbf{y}}}
\def\fatY{\mathbf{Y}}
\def\fatz{\mathbf{z}}
\def\fatZ{\mathbf{Z}}
\def\fat0{\mathbf{0}}

\def\tfatgamma{\tilde{\boldsymbol{\gamma}}}

\def\fatLambda{\boldsymbol{\Lambda}}
\def\fatmu{\boldsymbol{\mu}}

\def\fattheta{\boldsymbol{\theta}}

\def\tM{\tilde{M}}
\def\tS{\tilde{S}}
\def\Ktimes{\otimes}
\def\by{\bar{y}}
\def\bgamma{\bar{\gamma}}
\def\hgamma{\hat{\gamma}}
\def\tgamma{\tilde{\gamma}}

\def\pomega{\partial\omega}
\def\pOmega{\partial\Omega}

\def\Dxi{\Delta\xi}

\def\fatoy{\overline{\faty}}

\def\mN{\mathbb{N}}
\def\mR{\mathbb{R}}

\title{Mesoscopic Modeling of Random Walk and Reactions in
  Crowded Media}

\author[1]{Stefan Engblom}
\author[1]{Per L{\"o}tstedt}
\author[2]{Lina Meinecke}

\affil[1]{{\footnotesize Division of Scientific Computing, Department
    of Information Technology, Uppsala University, SE-751 05 Uppsala,
    Sweden. E-mail: \href{mailto:stefane@it.uu.se}{stefane},
    \href{mailto:perl@it.uu.se}{perl@it.uu.se}}}

\affil[2]{{\footnotesize Department of Mathematics, University of
    California, Irvine, CA 92697-3875, USA.  E-mail:
    \href{mailto:lina.meinecke@uci.edu}{lina.meinecke@uci.edu}}}

\begin{document}

\selectlanguage{english}
\maketitle

\begin{abstract}
  We develop a mesoscopic modeling framework for diffusion in a
  crowded environment, particularly targeting applications in the
  modeling of living cells. Through homogenization techniques we
  effectively coarse-grain a detailed microscopic description into a
  previously developed internal state diffusive framework. The
  observables in the mesoscopic model correspond to solutions of
  macroscopic partial differential equations driven by stochastically
  varying diffusion fields in space and time. Analytical solutions and
  numerical experiments illustrate the framework.
\end{abstract}

{\bf Keywords}: subdiffusion, crowding, internal states,
reaction-diffusion system.




\section{Introduction}

Living cells are controlled by a complicated network of
reaction-diffusion events. An example is exogenous signals triggering
the cell's response by reacting with the proteins present in the cell
or binding to the DNA to initiate transcription of certain genes. An
important task in computational systems biology is to study these
processes as accurately as possible inside the complicated cell
geometry. We specifically target two special features in a model of
the biochemical processes in living cells in this article: \emph{the
  high percentage of occupied volume in the cytoplasm} and \emph{the
  intrinsic noise}.

\begin{SCfigure}
  \centering
  \includegraphics[width = 0.4\textwidth]{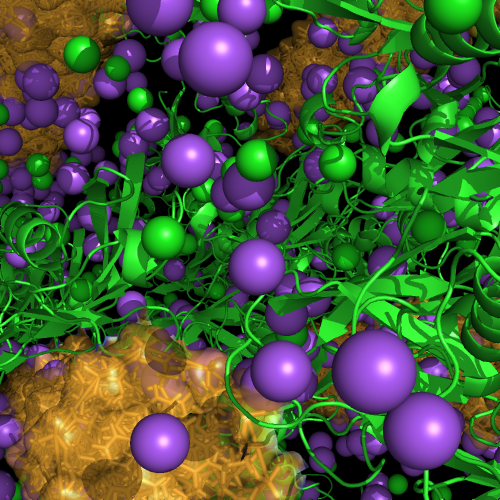}
  \caption{A snapshot from a molecular dynamics simulation: the
    interior of an \textit{E.~coli} is a highly crowded
    environment. Picture courtesy of David van der Spoel, Uppsala
    university.}
  \label{fig:coli}
\end{SCfigure}

It is estimated that up to 40\% of the available space in the
cytoplasm is occupied by macromolecules \cite{Luby-Phelps00,
  Schnell2004} and these have been shown to alter the dynamics of the
reaction network \cite{WoMu14}, see also Figure~\ref{fig:coli}.  Due
to the multiple steric repulsions between the tracer molecules and the
crowders, diffusion is slowed down. This macromolecular crowding
effect plays an even more important role on the cell membrane
\cite{Grasberger86}, where actin filaments create barriers for the
motion of membrane bound proteins \cite{Jin2007, Krapf15,
  Medalia2002}.

New imaging techniques \cite{DiRienzo2014} have shown that the
slowdown happens gradually over time. The tracers initially diffuse
freely without encountering the crowding macromolecules. Then they
start colliding with the crowders and go through an anomalous phase of
diffusion where their movement is constantly slowed down and their
mean square displacement (MSD) therefore grows sublinearly with time
\cite{Galanti14, HofFra, NiHaBu07}. On a long time scale they can be
observed to be diffusing normally again, but at a reduced diffusion
constant compared to the tracer in a dilute medium. This change in
diffusivity is a hydrodynamic consequence of the highly crowded space
inside cells. Moreover, the crowders also exhibit a thermodynamic
effect on the chemical reactions \cite{Hall2003}, which can be both
impeded (due to the longer time until collision) or facilitated (due
to the smaller effective reaction volume). In this paper, we will only
investigate inert crowders and their effect due to steric repulsion
with the reacting molecules. More complicated interactions such as
transient binding or interaction potentials further impact the
reaction-diffusion dynamics in a crowded environment \cite{Ando10,
  Penington2011, Saxton07, YAL04}, but lie outside the scope of this
study.

The second feature we incorporate in our modeling framework is
stochasticity. Although the cytosol is densely packed with molecules,
the individual species is often present at low copy numbers. A
deterministic macroscopic model describing the mean value of the
concentrations of the chemical species is therefore not applicable and
stochastic models remain as the computationally feasible alternative
\cite{KiNgWaBoSuTa14, McAArk97, PedOud05, ShaSwa08, SwaElo02}. On a
mesoscopic or on-lattice level of modeling, the domain is partitioned
into voxels and diffusion is modeled as a random jump process of the
molecules between the voxels. Inside each voxel, space is not resolved
further and the molecules are assumed to be well mixed and react
randomly with other molecules located within the same voxel. The time
evolution of this system is described by the
reaction-diffusion master equation \cite{VanKampen}. We sample
trajectories of the system using stochastic simulation techniques as
popularized by Gillespie \cite{gillespie}, originally developed for
well stirred problems without spatial dependence. Discretizations of
spatial domains were first considered in \cite{Elf2004, mesoRD, IsP}
and later improved to allow for unstructured meshes in
\cite{URDMEpaper, EnFeHeLo}. Overviews of deterministic, macroscopic
and stochastic, mesoscopic and microscopic levels of modeling of
biochemical networks are found in e.g. \cite{Burr17, EngHelLot17,
  MahmutovicElf}.

There have been several models combining the macromolecular crowding
effects and the stochastic mesoscopic level. In \cite{Roberts2013} the
most highly crowded voxels are defined as full and are made
inaccessible for the tracer molecules in order to model crowding. A
more gradual approach is to define the number of possible molecules
per voxel and then rescale the propensity to jump into this voxel by
how many spots are already occupied by other molecules
\cite{Fanelli10, Landman11, Taylor2015}. But by averaging the effect
of the crowders over the whole voxel, the transient anomalous phase is
not captured and we only observe the long time slower diffusion. To
resolve the short-time microscopic information the positions of
stationary obstacles were homogenized (or coarse-grained, upscaled) to
mesoscopic jump rates in \cite{Meinecke16}. Here, the crowders can
have arbitrary shape, but the diffusing tracers are understood to be
circular in two space dimensions (2D) and spherical in three
dimensions (3D).

In Brownian dynamics each individual molecule is tracked in a
lattice-free (or off-lattice) microscopic model. Here, all molecules
are spherical, move in Brownian motion, and react with a certain
probability when they touch each other \cite{Andrews10, ZoWo5a}.
Crowding is automatically incorporated in the model by the excluded
volume of the stationary or moving crowders. A stochastic, microscopic
simulation is in general more accurate than a mesoscopic simulation
but also much more computationally expensive. Microscopic simulation
of crowding and diffusion at the particle level is proposed in
\cite{SmithGrima17} and is evaluated in \cite{MARQUEZLAGO12,
  NiHaBu07}. In \cite{MeEr16} off-grid microscopic simulations are
compared to grid based microscopic cellular automata simulations and
the grid artifacts are quantified.

On the deterministic, macroscopic level, anomalous diffusion of the
concentrations due to crowding can be modeled by fractional partial
differential equations (FPDEs) \cite{BaGaMe12, HofFra, METZLER00}.
Internal states are introduced in \cite{MOMMER09} on the mesoscopic
level to model anomalous diffusion and in \cite{BEHL16} for reactions.
The internal state of a molecule changes with a certain probability
and determines the molecule's diffusion speed. The intensities for
these changes are given by the macroscopic FPDE for the observed
variables in \cite{BEHL16, MOMMER09}. Memory effects are included
without sacrificing the Markov property using these internal states.
Three physical interpretations of these internal states are that the
molecule is in different geometrical conformations, has different
methylation or phosphorylation, or resides in differently crowded
environments, which are all affecting the diffusion speed and reaction
propensities. Hidden states are also introduced in \cite{PLUE13} to
explain data from single cell experiments.

In this paper, we will combine the internal states model derived in
\cite{BEHL16} with the multiscale approach in \cite{Meinecke16} to
efficiently model diffusion of tracer particles among stationary or
moving crowder obstacles. The method
\begin{enumerate}
\item is considerably faster than Brownian dynamics,
\item allows more versatile modeling than mesoscopic methods where a
  limited number of molecules can occupy a lattice node,
\item defines a random diffusion field for a macroscopic
  equation expressed in observables.
\end{enumerate}

We first coarse-grain the microscale to the mesoscale by determining
statistics for the variation in the diffusion coefficient with the
homogenization method in \cite{Meinecke16}. The parameters of the
internal states model in \cite{BEHL16} can subsequently be deduced
from these data. Our mesoscopic method for crowding is less heuristic
than other methods and can be defined by experimental data, e.g., from
\cite{PLUE13}. The mesoscale equations are coarse-grained to
the macroscale analytically resulting in PDEs for the observables.


In the next section we first present the two mesoscopic models from
\cite{BEHL16} and \cite{Meinecke16} in more detail. We couple the
statistics from the microlevel to the parameters in the internal state
model in Section~\ref{sec:Combine}. The distributions of the molecules
in certain chemical systems with internal states and diffusion are
multinomial as shown by the analysis in Section~\ref{sec:analyt}. In
Section~\ref{sec:examples}, we test the resulting coarse-grained model
in examples in 2D and 3D and a summarizing discussion is found in the
final section.


\section{Two mesoscopic models}
\label{sec:two_meso}

The effect of static crowding molecules is coarse-grained from
the microscopic to the mesoscopic level of approximation according to
\cite{Meinecke16}. Then a discretized mesoscopic model built from an
internal states approximation is reviewed following \cite{BEHL16,
  MOMMER09}.

\subsection{Microscopic to mesosocopic model via first exit times}
\label{sec:multiscale}

\begin{figure}[H]
  \subfigure[]{\includegraphics[height=.16\textheight]{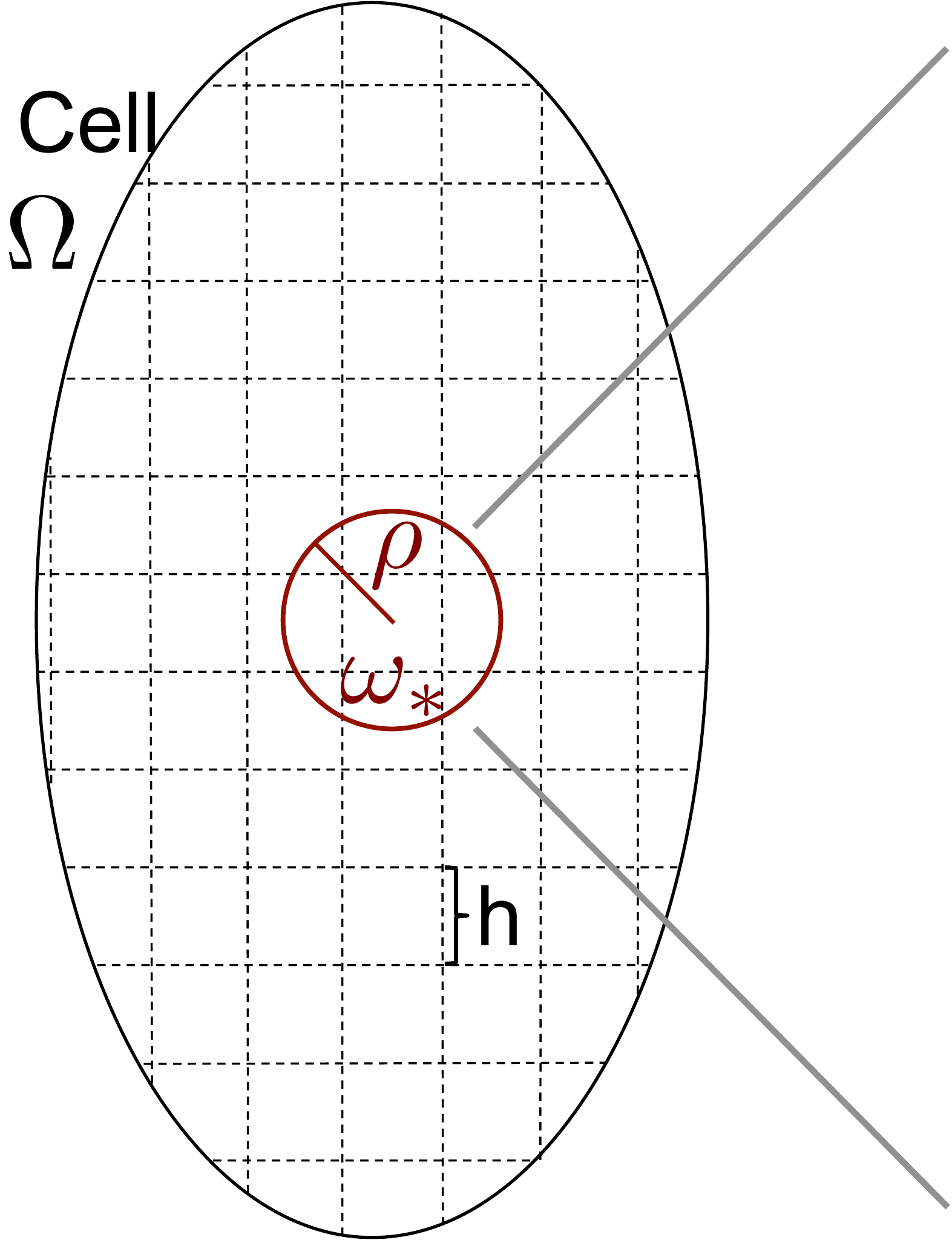}}
  \subfigure[]{\includegraphics[height=.16\textheight]{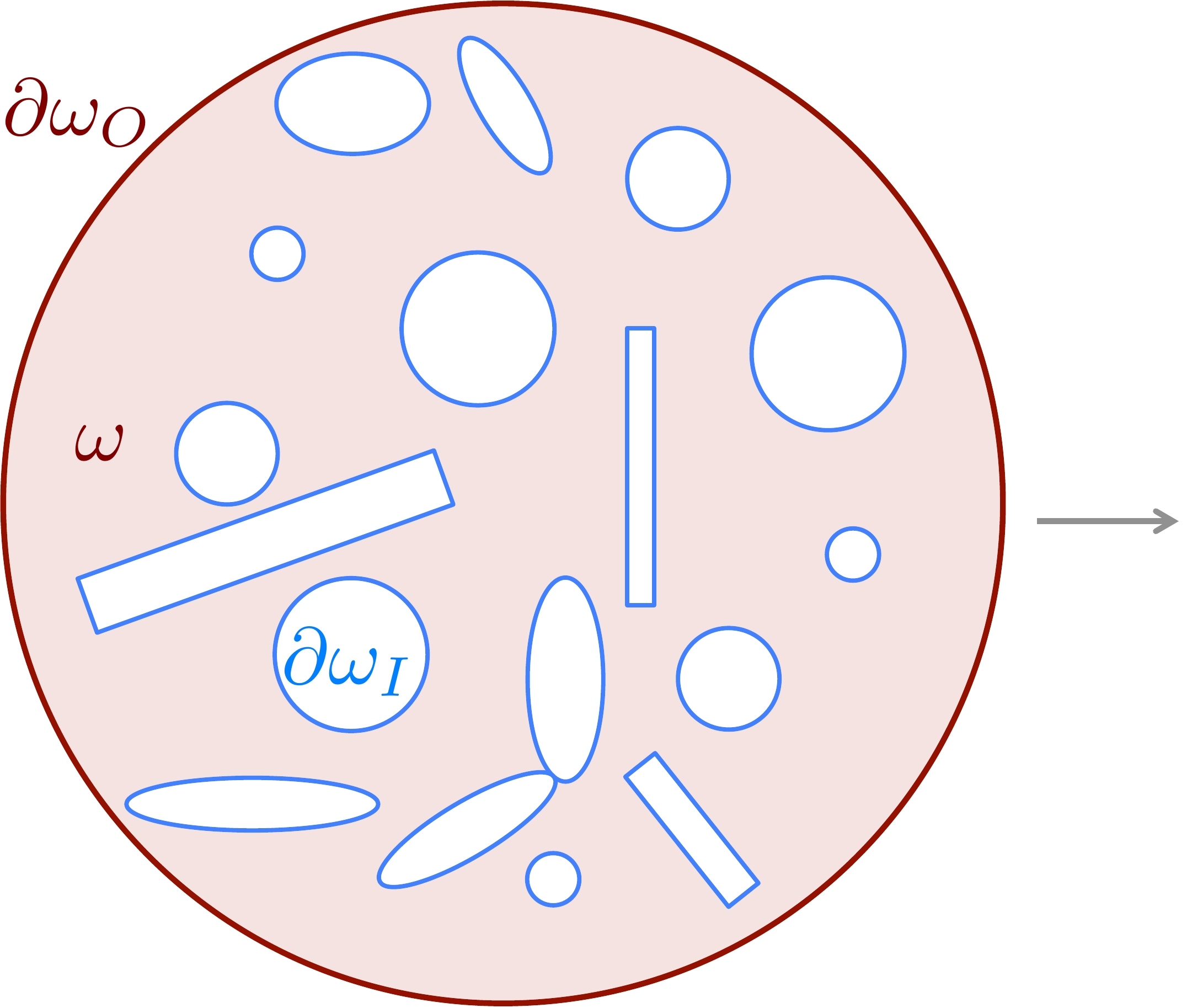}}
  \subfigure[]{\includegraphics[height=.16\textheight]{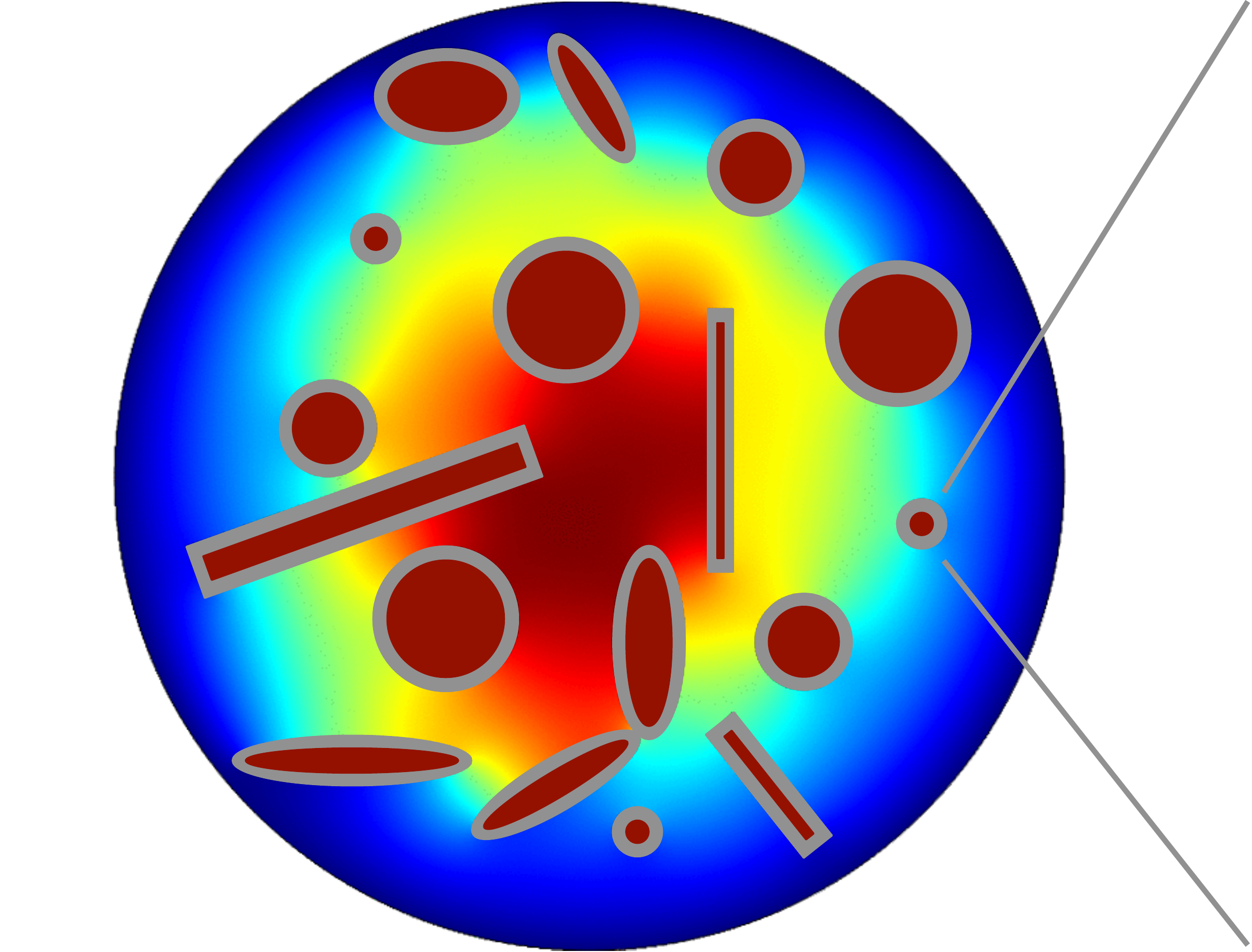}}
  \subfigure[]{\includegraphics[height=.16\textheight]{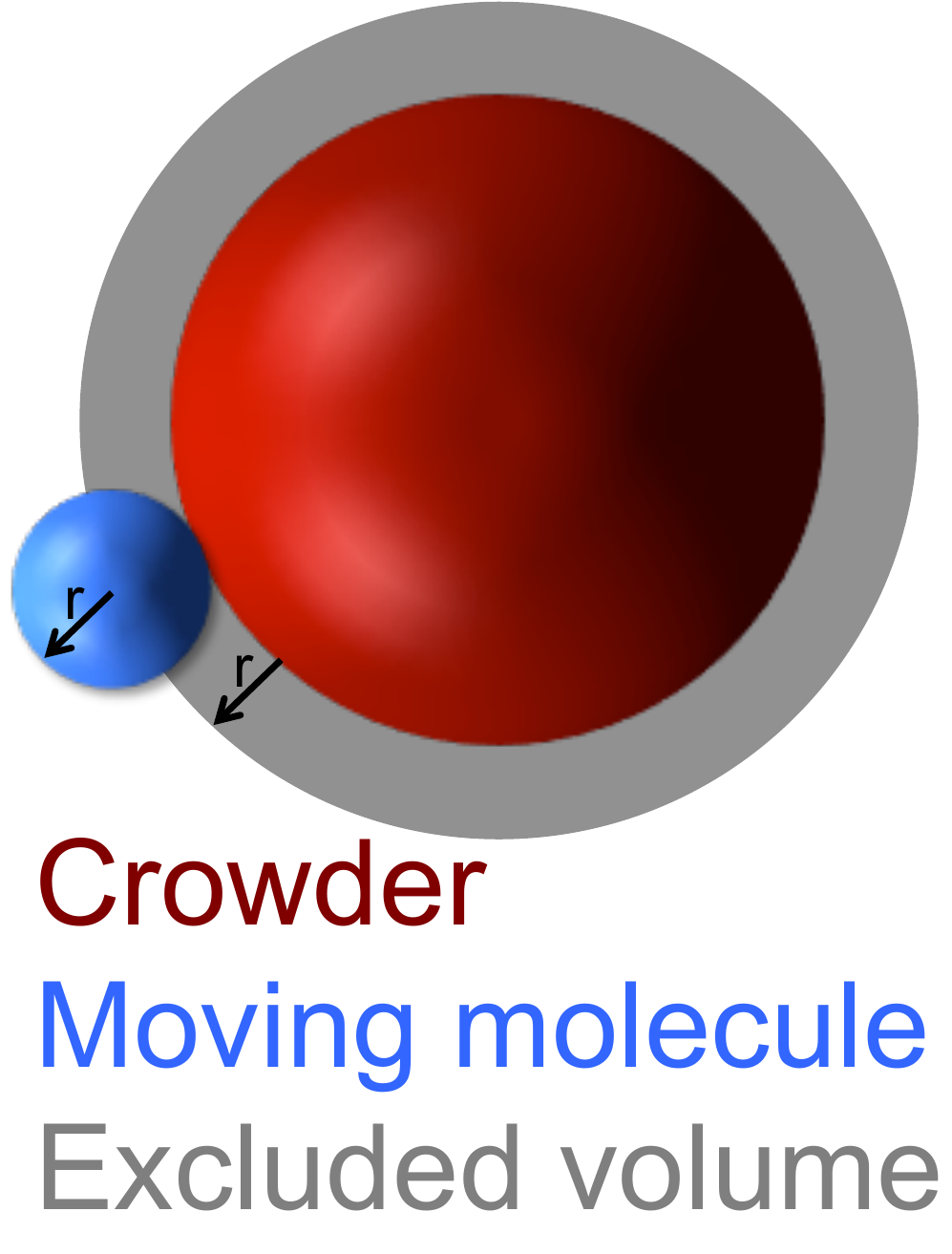}}
  \caption{(a) The cell volume $\Omega$ discretized by a grid with
    size $h$ giving rise to the non-perforated subvolumes
    $\omega_{\ast}$. (b) The circular, perforated domain $\omega$ (pink) of
    radius $\rho$ where the cut-outs represent the obstacles, and the
    outer $\partial\omega_O$ and inner $\partial\omega_I$
    boundaries. (c) Solution to \eqref{eq:Poiss1} on $\omega$ with
    crowders represented as holes with reflective boundary conditions
    and with high values of $E(\fatx)$ in red and low values in
    blue. (d) The excluded volume consists of the volume occupied by
    the crowding molecule enlarged by the radius $r$ of the diffusing
    tracer molecule.}
  \label{fig:omega}
\end{figure}

Single tracer molecules move on the microscopic scale by Brownian
motion in a domain $\omega_\ast$ with obstacles. The moving molecules
are assumed to be circular in 2D and spherical in 3D with radius $r$.
The crowder obstacles are stationary in space and chemically inert,
such that the boundary condition for the moving molecule is reflective
at the surface of the crowding objects, which are represented
explicitly as holes in $\omega_\ast$.  The volume of the interior of
the cell is denoted by $\Omega$ and $\omega_\ast$ is a subvolume of
$\Omega$, $\omega_\ast\subset\Omega$.  The subvolumes $\omega_{\ast}$
are occupied by crowders such that the free space remaining for the
moving molecule is $\omega$ and $\omega\subseteq\omega_\ast$, see
Figure~\ref{fig:omega}(a) and (b).

Let $\gamma_0$ be the diffusion coefficient for the Brownian motion.
In \cite{Meinecke16} we presented a multiscale approach to compute the
effective diffusion rate $\gamma$ in the crowded environment
$\omega_\ast$ using the mean value of the first exit time $E(\fatx)$,
see \cite{Oksendal}, from $\omega$ fulfilling
\begin{align}
  \gamma_0\Delta E(\fatx)&=-1, &\fatx&\in\omega,\label{eq:Poiss1}\\
  E(\fatx) &=0, &\fatx&\in\partial\omega_O,\label{eq:Poiss2}\\
  \fatn\cdot\nabla E(\fatx)&=0,&\fatx&\in\partial\omega_I.\label{eq:Poiss3}
\end{align}
The starting position of the diffusing molecule is $\fatx\in\omega$,
$\partial\omega_O$ is the outer boundary of $\omega$ shared with
$\omega_\ast$ and $\partial\omega_I$ is the inner boundary of the
obstacles with normal $\fatn$, see Figure~\ref{fig:omega}(b).  Since
\eqref{eq:Poiss1} describes the expected exit time of a moving point
particle, the cut-outs in the perforated domain are enlarged to
account for the radius $r$ of the tracer, see
Figure~\ref{fig:omega}(d).

Equation \eqref{eq:Poiss1} and boundary condition \eqref{eq:Poiss2}
also hold on the non-perforated domain $\omega_\ast$ without the
boundary condition on $\partial\omega_I$ and resulting in the solution
$E_\ast(\fatx)$.  Since the mean first exit time in \eqref{eq:Poiss1}
is inversely proportional to the diffusion coefficient it can be used
to compute the effective diffusion rate $\gamma$ in the crowded domain
$\omega_\ast$ according to
\begin{equation}
  \gamma=\gamma_0\frac{E_\ast(\fatx)}{E(\fatx)}.
\label{eq:gammaform}
\end{equation}
In this way, all the details in $\omega$ are avoided and an effective
(or homogenized, upscaled, coarse-grained) diffusion coefficient
$\gamma$ in $\omega_\ast$ is determined.  The domain $\omega_\ast$ and
point $\fatx$ are arbitrary but if $\omega_\ast$ is circular ($d=2$)
or spherical ($d=3$) with radius $\rho$ then (see \cite{Oksendal})
\begin{equation}
  E_\ast(\fatx)=\frac{1}{\gamma_0}\frac{\rho^2-\|\fatx\|^2_2}{2d}.
  \label{eq:Esol}
\end{equation}
The effective diffusion rate $\gamma$ depends on the percentage of
excluded volume $\phi\in[0,1]$. If $\phi=0$ then
$E_\ast(\fatx)=E(\fatx)$ and $\gamma=\gamma_0$ in \eqref{eq:gammaform}
and if $\phi\rightarrow 1$ then there is no space left for molecular
motion, $E(\fatx)\rightarrow \infty$, and $\gamma\rightarrow
0$. Depending on the shape of the obstacles and the size of the moving
molecule $r$, $\gamma$ can be $0$ for a small $\phi<1$.

This approach is universal in the way that the stationary crowding
molecules can have any shape. New $\gamma$ values for other shapes are
determined in \cite{Meinecke16}.  If the crowders are spherical with
radius $R$, the radii have to satisfy $r,R\ll \rho$ for the upscaling
to be accurate and not too sensitive to the particular distribution of
the obstacles. The typical size of a moving or crowding molecule is
4-20 nm (globular protein-ribosome). Then a possible $\rho$ is
$\rho\sim 50$ nm, which is sufficiently small to discretize a
prokaryote {\it E.~coli} of size $1-3\,\mu {\rm m}$ or an eukaryote
cell which is about ten times larger.

\subsection{The internal states model}
\label{sec:intstates}

The mesoscopic model for diffusion and chemical reactions is extended
such that each molecule can adopt several internal states that may be
unobservable. These extra internal states are used to model
subdiffusion in \cite{BEHL16, MOMMER09} and will be used here to
represent crowding with moving obstacles.

\subsubsection{The spatial internal states model}
\label{sec:intstmodel}

Let $u(\fatx, t, \xi)$ be the concentration of a molecular species at
$\fatx\in\Omega$ at time $t\ge 0$ in a continuous internal state
$\xi\in\Xi=[0, \xi_{\max}] \subset \mR_+$. The rate of change from
internal state $\eta$ to state $\xi$ is $A(\xi, \eta)$. At the
boundary $\partial\Omega$ of $\Omega$ the molecules are reflected and
the scalar diffusion $\gamma(\xi)$ is allowed to depend on the
internal state. Then $u$ satisfies, for $t>0$,
\begin{equation}
\begin{array}{rl}
  u_{t}(\fatx, t, \xi)&=\nabla\cdot\left( \gamma(\xi)\nabla u(\fatx, t, \xi)\right)+\int_{\Xi} A(\xi, \eta)u(\fatx, t, \eta)\,d\eta,\quad \fatx\in\Omega, \; \xi\in\Xi,\\
   \fatn\cdot\nabla u&=0,\quad \fatx\in\pOmega,\; \xi\in\Xi.
\end{array}
\label{eq:ueq}
\end{equation}

The total amount of the species $\int_{\Omega} \int_{\Xi} u(\fatx, t,
\xi)\,d\xi\,d\fatx$ should remain constant for mass conservation. By
integrating \eqref{eq:ueq} over $\Xi$ and $\Omega$ using the boundary
condition on $\pOmega$, we obtain the time derivative
\begin{equation}
  \begin{array}{lcl}
  \displaystyle{\partial_t \int_{\Xi}\int_\Omega u(\fatx, t, \xi)\,d\xi\, d\fatx}\\
=\displaystyle{\int_{\Xi}\int_\Omega\nabla\cdot \left(\gamma(\xi)\nabla u(\fatx, t, \xi)\right)\, d\fatx\, d\xi}
   \displaystyle{+\int_{\Xi}\int_\Omega\int_{\Xi} A(\xi, \eta)u(\fatx, t, \eta)\,d\eta\, d\fatx\, d\xi}\\
  =\displaystyle{\int_{\Xi} \gamma(\xi)\int_{\pOmega}\fatn\cdot\nabla u(\fatx, t, \xi)\, ds\, d\xi}
   \displaystyle{ + \int_\Omega\int_{\Xi} u(\fatx, t, \eta)\int_{\Xi} A(\xi, \eta)\, d\xi\, d\eta \, d\fatx}\\
  =\displaystyle{\int_\Omega \int_{\Xi}u(\fatx, t, \eta) \int_{\Xi}A(\xi, \eta)\, d\xi\, d\eta\, d\fatx}.
  \end{array}
\label{eq:ueqint}
\end{equation}
The time derivative of the total amount must vanish for all $u$. Thus,
a sufficient condition on $A$ for this to hold is
\begin{equation}
  \int_{\Xi} A(\xi, \eta)\, d\xi=0.
\label{eq:Acond}
\end{equation}
The observable $U(\fatx,t)$ denotes the concentration of the molecule
independent of its internal state and is defined by
\begin{equation}
  U(\fatx, t)= \int_{\Xi} u(\fatx, t, \xi)\, d\xi.
\label{eq:Udef}
\end{equation}
If $A$ is chosen as in \eqref{eq:Acond} then by \eqref{eq:ueq}, $U$ in
\eqref{eq:Udef} satisfies
\begin{equation}
  U_t(\fatx, t)=\nabla\cdot\int_{\Xi} \gamma(\xi)\nabla u(\fatx, t, \xi)\, d\xi.
\label{eq:Ueq1}
\end{equation}
There is an ordinary diffusion equation for $U$ only if $\gamma$ is
independent of $\xi$. With a diffusion tensor $\tgamma_i(\fatx, t), \;
i=1,\ldots,d,$ such that
\[
\tgamma_i(\fatx, t)=\int_{\Xi} \gamma(\xi)\partial_{x_i} u(\fatx, t, \xi)\, d\xi/\partial_{x_i} \int_{\Xi} u(\fatx, t, \xi)\, d\xi,
\]
\eqref{eq:Ueq1} can be written
\begin{equation}
  U_t=\nabla\cdot\left(\tfatgamma(\fatx, t)\nabla U\right),
\label{eq:Ueq2}
\end{equation}
but in general $\tfatgamma$ is not known explicitly.

A particular choice of $A$ is
\begin{equation}
  A(\xi, \eta)=(\mu(\xi)-\delta(\xi-\eta))T(\eta)
\label{eq:Aspec}
\end{equation}
with the Dirac measure $\delta$. Then \eqref{eq:Acond} is fulfilled if
$\mu$ is scaled such that
\begin{equation}
  \int_{\Xi}\mu(\xi)\, d\xi=1.
\label{eq:muscal}
\end{equation}

In order to find the steady state solution of \eqref{eq:ueq}, we set tentatively
\begin{equation}
  u_\infty(\xi)=\mu(\xi)/T(\xi).
\label{eq:uinf}
\end{equation}
Clearly,
\begin{equation}
  \int_{\Xi} A(\xi, \eta)u_\infty(\eta)\, d\eta=0,
\label{eq:Anull}
\end{equation}
and $u_\infty$ is indeed a steady state solution of
\eqref{eq:ueq}. A natural convention is to let $\mu(\xi)\ge 0$ and
$T(\xi)>0$ for $u_\infty$ to be a non-negative concentration.

\subsubsection{Discretization in space and internal states}
\label{sec:discrdiff}
 
Let $\Omega$ be discretized by a triangular (2D) or tetrahedral (3D)
primal mesh with nodes at $\fatx_i,\, i=1,\ldots, J$.  The dual mesh
consists of voxels $\calV_i,\, i=1,\ldots,J$, as in
\cite{EnFeHeLo}. Each node $\fatx_i$ is associated with one voxel
$\calV_i$.  The solution $u$ of \eqref{eq:ueq} is approximated by the
finite element method using linear basis functions $\varphi_i(\fatx)$
satisfying $\varphi_i(\fatx_i)=1$ and $\varphi_i(\fatx_j)=0,\, j\ne
i$.  The internal state space $\Xi$ is partitioned into $K$ intervals
$\calI_k=[\xi_{k-1}, \xi_k],\, k=1,\ldots,K,$ of length
$\Dxi=\xi_{\max}/K$.  In each interval $k$ there is a basis function
$\psi_k$ such that $\psi_k(\xi)=1/\sqrt{\Dxi},\, \xi\in\calI_k$, and
$\psi_k(\xi)=0$ otherwise. Then $u_h$ approximating $u$ is
\begin{equation}
  u_h(\fatx, t, \xi)=
  \sum_{j=1}^J\sum_{l=1}^K u_{jl}(t)\varphi_j(\fatx)\psi_{l}(\xi).
  \label{eq:uhdef}
\end{equation}
Insert $u_h$ into \eqref{eq:ueq}, multiply by a test function
$\varphi_i(\fatx)\psi_{k}(\xi)$ in a tensor product finite element
space and integrate over $\Omega\times\Xi$ to derive an equation for
the evolution of $u_{jl}$. If $\xi$ in \eqref{eq:ueq} is interpreted
as a random variable determining the diffusion, then \eqref{eq:uhdef}
is the approximation suitable for a stochastic Galerkin method to
solve \eqref{eq:ueq} \cite{GuWeZh14}.

Let $E_\alpha$ be a triangular element in 2D or a tetrahedral element in 3D with area or
volume $|E_\alpha|$ and $\calT_{ij}$ the set of triangles or
tetrahedra with a common edge $ij$ between nodes $i$ and $j$.  The
Kronecker delta is denoted by $\delta_{ij}$. An element in the
stiffness tensor $S$ is then
\begin{equation}
\begin{array}{rl}
  S_{ijkl}&=\displaystyle{-\sum_{E_\alpha\in\calT_{ij}}\int_{E_\alpha} \nabla\varphi_i^T|_{E_\alpha} \nabla\varphi_j|_{E_\alpha}\, d\fatx \int_{\Xi}\gamma(\xi)\psi_{k}(\xi)\psi_{l}(\xi)\, d\xi}\\
        &=\displaystyle{-\sum_{E_\alpha\in\calT_{ij}}\nabla\varphi_i^T|_{E_\alpha} \nabla\varphi_j|_{E_\alpha}|E_\alpha|\gamma_k\delta_{kl}=\gamma_kS_{ij}\delta_{kl},}
\end{array}
\label{eq:Sijkl}
\end{equation}
where $\gamma_k$ is the average of $\gamma(\xi)$ in $\calI_k$.  In the
diagonal element with $i=j$, the integration domain in $\fatx$ is over
all $E_\alpha$ with a corner at $\fatx_i$.  Choose $\gamma_k$ to be
$\gamma_0 T(k)$ and let $\fatT$ be the matrix with $T(k)$ in the
diagonal. Hence, with a stiffness matrix
$\fatS$ the stiffness tensor in \eqref{eq:Sijkl} can be written
\begin{equation}
  S=\gamma_0\fatS\Ktimes\fatT,
  \label{eq:SijTk}
\end{equation}
where $\Ktimes$ denotes the Kronecker product.  

The mass tensor $\tM$ is defined by
\begin{equation}
  \tM_{ijkl}=\sum_{E_\alpha\in\calT_{ij}}\int_{E_\alpha} \varphi_i(\fatx)\varphi_j(\fatx)\, d\fatx\,\int_{\Xi} \psi_k(\xi)\psi_l(\xi)\, d\xi= \tM_{ij}\delta_{kl}.
\label{eq:Mijkl}
\end{equation}
The first part of $\tilde{\fatM}$ depends on the geometry and is
lumped and replaced by a diagonal matrix $\fatM$ such that
\begin{equation}
  M_{ij}=M_i\delta_{ij},\quad M_i=\sum_{l=1}^J \tM_{il}.
\label{eq:Mijlump}
\end{equation}
Then by \eqref{eq:Mijkl} 
\begin{equation}
  \tM_{ijkl}=M_i\delta_{ij}\delta_{kl}.
\label{eq:Mij}
\end{equation}

An element in the tensor discretizing the operator $A$ for change of
internal state is
\begin{equation}
\begin{array}{ll}
  A_{ijkl}&=\displaystyle{-\sum_{E_\alpha\in\calT_{ij}}\int_{E_\alpha} \varphi_i(\fatx)\varphi_j(\fatx)\, d\fatx \int_\Xi \int_\Xi A(\xi, \eta)\psi_{k}(\xi)\psi_{l}(\eta)\, d\xi\, d\eta}\\
        &=\displaystyle{\kappa_0\tM_{ij}A_{kl}},
\end{array}
\label{eq:Aijkl}
\end{equation}
where $A_{kl}$ is an element in the matrix $\fatA$ and $\kappa_0$ is a
freely choosable scaling of $\fatA$ that denotes how fast the
molecules change their internal state. Thus, $A$ can be written as
$A=\kappa_0\fatM\Ktimes\fatA$ after mass lumping of $\tM$.

Let $\fate_J$ be defined by $\fate^T_J=(1,1,\ldots,1)\in\mR^J$ and let
$\mu(\xi)$ be a piecewise constant function such that
$\mu(\xi)=\sum_{k=1}^K \mu_k\psi_k(\xi)$. Then the matrix-vector forms
of the condition in \eqref{eq:Acond}, the special choice of $A$ in
\eqref{eq:Aspec}, the scaling of the components of $\fatmu$ in
\eqref{eq:muscal}, and the null vector $\fatu_{i\infty}$ of $\fatA$ in
\eqref{eq:uinf} are
\begin{equation}
\begin{array}{ll}
  \fate_K^T\fatA=\fat0,\; 
  \fatA=(\fatmu\fate_K^T-\fatI_K)\fatT, \; \fate_K^T\fatmu=1,\; \\
  \fatu_{i\infty}=\fatT^{-1}\fatmu \Longrightarrow \fatA\fatu_{i\infty}=\fat0,
\end{array}
\label{eq:Aprop}
\end{equation}
where $\fatI_J$ is the identity matrix of dimension $J\times J$. These
properties are shared by $\fatA$ in \cite{BEHL16, MOMMER09}. Since
$\fate_K^T\fatA=\fat0$ there is one eigenvalue of $\fatA$ equal to $0$
with eigenvector $\fatu_{i\infty}$. The diagonal elements of $\fatA$
are negative and it follows from Gerschgorin's theorem that the real
parts of the eigenvalues of $\fatA$ are non-positive.

The diffusion matrix $\fatD$ is defined by
\begin{equation}
  \fatD=\gamma_0\fatM^{-1}\fatS.
\label{eq:Ddef}
\end{equation}
With the expressions derived in \eqref{eq:SijTk} and \eqref{eq:Aijkl}
and multiplication by the inverse of the lumped mass matrix, the
discretized equation \eqref{eq:ueq} for all concentrations $\fatu$ is
\begin{equation}
  \fatu_t=\gamma_0(\fatD\Ktimes\fatT)\fatu+\kappa_0(\fatI_J\Ktimes\fatA)\fatu,
\label{eq:diffequ1}
\end{equation}
or for the concentration $\fatu_i$ in voxel $i$
\begin{equation}
  \fatu_{it}=\gamma_0\fatT\left(\sum_{j\in\calJ(i)}D_{ij}\fatu_j+D_{ii}\fatu_i\right)+\kappa_0\fatA\fatu_i,\; i=1,\ldots,J.
\label{eq:diffequ2}
\end{equation}
The index set $\calJ(i)$ consists of the indices $j$ with an edge
connecting $\fatx_i$ and $\fatx_j$ implying that $D_{ij}\ne 0$. The vector
$\fatu\in\mR^{JK}$ has components $u_{ik},\; i=1,2,\ldots,J,\;
k=1,2,\ldots,K,$ denoting the concentration in the internal state $k$ at node or voxel $i$ and $\fatu_i$ is
a subset of $\fatu$ restricted to all the internal states in voxel $i$.

The mean values $\by_{ik}$ of the copy numbers of the species satisfy
\eqref{eq:diffequ2} with $\fatu_{i}=|\calV_i|^{-1}\fatby_{i}$
\begin{equation}
\begin{array}{rl}
  \fatby_{it}&=\displaystyle{\gamma_0\fatT\left(\sum_{j\in\calJ(i)}\frac{S_{ij}}{|\calV_j|}\fatby_j+\frac{S_{ii}}{|\calV_i|}\fatby_i\right)+\kappa_0\fatA\fatby_i}\\
            &=\displaystyle{\gamma_0\fatT\left(\sum_{j\in\calJ(i)}\lambda_{ji}\fatby_j-\lambda_{i}\fatby_i\right)+\kappa_0\fatA\fatby_i,\; i=1,\ldots,J,}
\end{array}
\label{eq:diffeqy1}
\end{equation}
where $\lambda_{ji}, S_{ij},$ and $D_{ij}$ in \eqref{eq:diffequ2} and \eqref{eq:diffeqy1} are related by
\begin{equation}
  \lambda_{ji}=\frac{S_{ij}}{|\calV_j|},\; D_{ij}=\frac{|\calV_j|}{|\calV_i|}\lambda_{ji},
     \; \lambda_{i}=-D_{ii},\;  \sum_{j\in\calJ(i)} |\calV_j|\lambda_{ji}=|\calV_i|\lambda_{i},\;  \sum_{i, i\ne j} \lambda_{ji}=\lambda_{j},
\label{eq:Dij}
\end{equation}
see \cite{EnFeHeLo}. The vector $\fatby$ holds $\fatby_i,\,
i=1,\ldots,J,$ stored consecutively. With $\Lambda_{ij}=\lambda_{ji}$,
the equation for $\fatby$ is similar to \eqref{eq:diffequ1},
\begin{equation}
  \fatby_t=\gamma_0(\fatLambda\Ktimes\fatT)\fatby+\kappa_0(\fatI_J\Ktimes\fatA)\fatby.
\label{eq:diffeqy2}
\end{equation}

The sum of the components in $\fatby$ is
\begin{equation}
  \sum_{i=1}^J\sum_{k=1}^K y_{ik}=(\fate_J\Ktimes\fate_K)^T\fatby.
\label{eq:sumy}
\end{equation}
By \eqref{eq:Dij} we have $\fate_J^T\fatLambda=\fat0$. Hence,
\begin{equation}
   ((\fate_J\Ktimes\fate_K)^T\fatby)_t=(\fate_J\otimes\fate_K)^T\fatby_t=\gamma_0(\fate_J^T\fatLambda\Ktimes\fate_K^T\fatT)\fatby+\kappa_0(\fate_J^T\fatI_J\Ktimes\fate_K^T\fatA)\fatby=\fat0,
\label{eq:diffsumy}
\end{equation}
since $\fate_K^T\fatA=\fat0$ in \eqref{eq:Aprop}. Consequently, the
sum in \eqref{eq:sumy} is constant in time
\begin{equation}
  \sum_{i=1}^J\sum_{k=1}^K y_{ik}(t)=\sum_{i=1}^J\sum_{k=1}^K y_{ik}(0),\quad t>0.
\label{eq:sumyconst}
\end{equation}
 
The jump coefficients $\lambda_{ji}\ge 0$ are proportional to the
probability of a molecule in voxel $\calV_j$ to jump to $\calV_i$ in a
stochastic simulation of the system \cite{EnFeHeLo}. A non-negative
$\lambda_{ji}$ is required for an interpretation of it as a
probability. In a mesh of poor quality, $\lambda_{ji}$ may be negative
due to an $S_{ij}<0$ in the finite element discretization but
corrections are derived in \cite{LoMe, MEHL15} such that
$\lambda_{ji}\ge 0$ on any mesh.

It follows from the properties of $\lambda_{ji}$ in \eqref{eq:Dij} and
$\fatA$ in \eqref{eq:Aprop} that there is a stationary solution
$\fatby_{i\infty}=v_i\fatT^{-1}\fatmu,\; i=1,\ldots,J,$ with
$v_i=|\calV_i|$ to \eqref{eq:diffeqy1} such that
\begin{equation}
  \fatby_{i\infty t}=\gamma_0\fatT\left(\sum_{j\in\calJ(i)}\lambda_{ji}\fatby_{j\infty}-\lambda_{i}\fatby_{i\infty}\right)+\kappa_0\fatA\fatby_{i\infty}=\fat0.
\label{eq:Duinf}
\end{equation}
Hence, with $\fatby_\infty=\by\fatv\Ktimes\fatT^{-1}\fatmu$ in \eqref{eq:diffeqy2} 
\begin{equation}
  \fatby_{\infty t}=\gamma_0\by\fatLambda\fatv\Ktimes\fatmu+\kappa_0\by\fatv\Ktimes\fatA\fatT^{-1}\fatmu=\fat0.
\label{eq:diffeqy2inf}
\end{equation}

The equation for the concentration observable $U_i=\fate_K^T\fatu_i$
in $\calV_i$, cf. \eqref{eq:Udef}, is by \eqref{eq:diffequ1} and
\eqref{eq:Aprop}
\begin{equation}
  U_{it}=\gamma_0\left(\sum_{j\in\calJ(i)}D_{ij}\fate_K^T\fatT\fatu_j+D_{ii}\fate_K^T\fatT\fatu_i\right).
\label{eq:diffeqUi}
\end{equation}
An explicit equation for $U_i$ is obtained if we knew the diffusion coefficient 
\begin{equation}
  \hgamma_j(t)=\gamma_0\fate_K^T\fatT\fatu_j/U_j  
\label{eq:hdiffcoeff}
\end{equation}
in $\calV_j$. Then by \eqref{eq:Ddef}, \eqref{eq:diffeqUi} is rewritten
\begin{equation}
  U_{it}=\sum_{j\in\calJ(i)} \hgamma_j(t)\frac{S_{ij}}{|\calV_i|}U_j+\hgamma_i(t)\frac{S_{ii}}{|\calV_i|}U_i.
\label{eq:diffeqUi2}
\end{equation}

The stiffness matrix with a variable diffusion in space and time in
\eqref{eq:Ueq2} is
\begin{equation}
\begin{array}{rl}
  \tS_{ij}(t)&=\displaystyle{-\sum_{E_\alpha\in\calT_{ij}}\int_{E_\alpha} \nabla\varphi_i^T|_{E_\alpha}\tgamma(\fatx, t) \nabla\varphi_j|_{E_\alpha}\, d\fatx}\\
        &=\displaystyle{-\sum_{E_\alpha\in\calT_{ij}}\nabla\varphi_i^T|_{E_\alpha} \nabla\varphi_j|_{E_\alpha}\tgamma_{\alpha}(t)|E_\alpha|=\tgamma_{ij}(t)S_{ij},}
\end{array}
\label{eq:tSij}
\end{equation}
where $\tgamma_{\alpha}$ is the spatial average of $\gamma(\fatx, t)$
in element $E_\alpha$ and the last equality defines $\tgamma_{ij}$ as in
\eqref{eq:Sijkl}.  Then, using $\tS_{ij}$ in \eqref{eq:tSij}, the
discretization of \eqref{eq:Ueq2} is
\begin{equation}
  U_{it}=\sum_{j\in\calI(i)} \frac{\tS_{ij}}{|\calV_i|}U_j+\frac{\tS_{ii}}{|\calV_i|}U_i
       =\sum_{j\in\calI(i)} \tgamma_{ij}(t)\frac{S_{ij}}{|\calV_i|}U_j+\tgamma_{ii}(t)\frac{S_{ii}}{|\calV_i|}U_i.
\label{eq:diffeqUi3}
\end{equation}
Thus, \eqref{eq:diffeqUi2} is a discretization of \eqref{eq:Ueq2} with
a time varying $\hgamma_j$. The diffusion coefficient $\tgamma_{ij}$
along the edges in the direct discretization of \eqref{eq:Ueq2} in
\eqref{eq:diffeqUi3} is approximated by $\hgamma_{j}$ at the nodes in
\eqref{eq:diffeqUi2}.
  
\subsubsection{Several species and reactions}
\label{sec:manyspecies}

Assume that the diffusion coefficient $\gamma_0$ and that the
transition matrix $\fatA$ are the same for all species in the
system. The components of the copy number vector $\fatby$ in
\eqref{eq:diffeqy1} are $\by_{ik\ell}$ where $\ell=1,2,\ldots,L,$
denotes the molecular species. The reactions are assumed to be the
same in every voxel independent of space, depending only on the copy
number in the voxel. They are also assumed to be the same in each
internal state except for a scaling with $\fatG$. In model I for the
reactions in \cite{BEHL16}, $\fatG=\fatT$, and in model II,
$\fatG=\fatI$. Then the reaction-diffusion equation is derived by
adding a reaction term $\fate_J\Ktimes\fatg\Ktimes\fatf$ with
$\fatg=\fatG\fate_K$ to \eqref{eq:diffequ1}, thus extending the
solution $\fatby$ in \eqref{eq:diffeqy1} by the number of molecules of
the different species. In the reaction term, $\fatf$ in the
$e_ig_k\fatf$ element of $\fate_J\Ktimes\fatg\Ktimes\fatf$ in voxel
$\calV_i$ and internal state $k$ depends on $\fatby_{ik}\in\mR^L$, the
state vector of copy numbers of the $L$ different species in $\calV_i$
in internal state $k$.  Including reactions, equation
\eqref{eq:diffeqy2inf} then becomes
\begin{equation}
  \fatby_t=\gamma_0(\fatLambda\Ktimes\fatT\Ktimes\fatI_L)\fatby+\kappa_0(\fatI_J\Ktimes\fatA\Ktimes\fatI_L)\fatby+\fate_J\Ktimes\fatg\Ktimes\fatf.
\label{eq:reacteq}
\end{equation}
Unless $\fatf$ is affine in $\fatby$, the solution of this macroscopic
reaction-diffusion equation only approximates the mean values of the
number of molecules in the mesoscopic model, see e.g. \cite{GaLeOt05}.
If $\fatT=\fatI$ in \eqref{eq:reacteq}, then the diffusion is the same
for the molecules in all internal states but may differ in the
reaction rates in $\fatf$.


\section{Connecting the multiscale and the internal states models}
\label{sec:Combine}

A constructive procedure to incorporate the coarse-grained diffusion
coefficients into the internal state framework is proposed in this
section. Briefly, the computed statistical distribution of the
crowding molecules is used to determine the parameters in the internal
state model.

\subsection{Coarse-graining the diffusion coefficient}
\label{sec:gamcomp}

If the obstacles are stationary and their shapes and positions are
known, then the effect of the crowding can be computed directly as in
Section~\ref{sec:multiscale} and there is no need for internal states.

\begin{figure}[H]
  \center{\includegraphics[width=.75\textwidth]{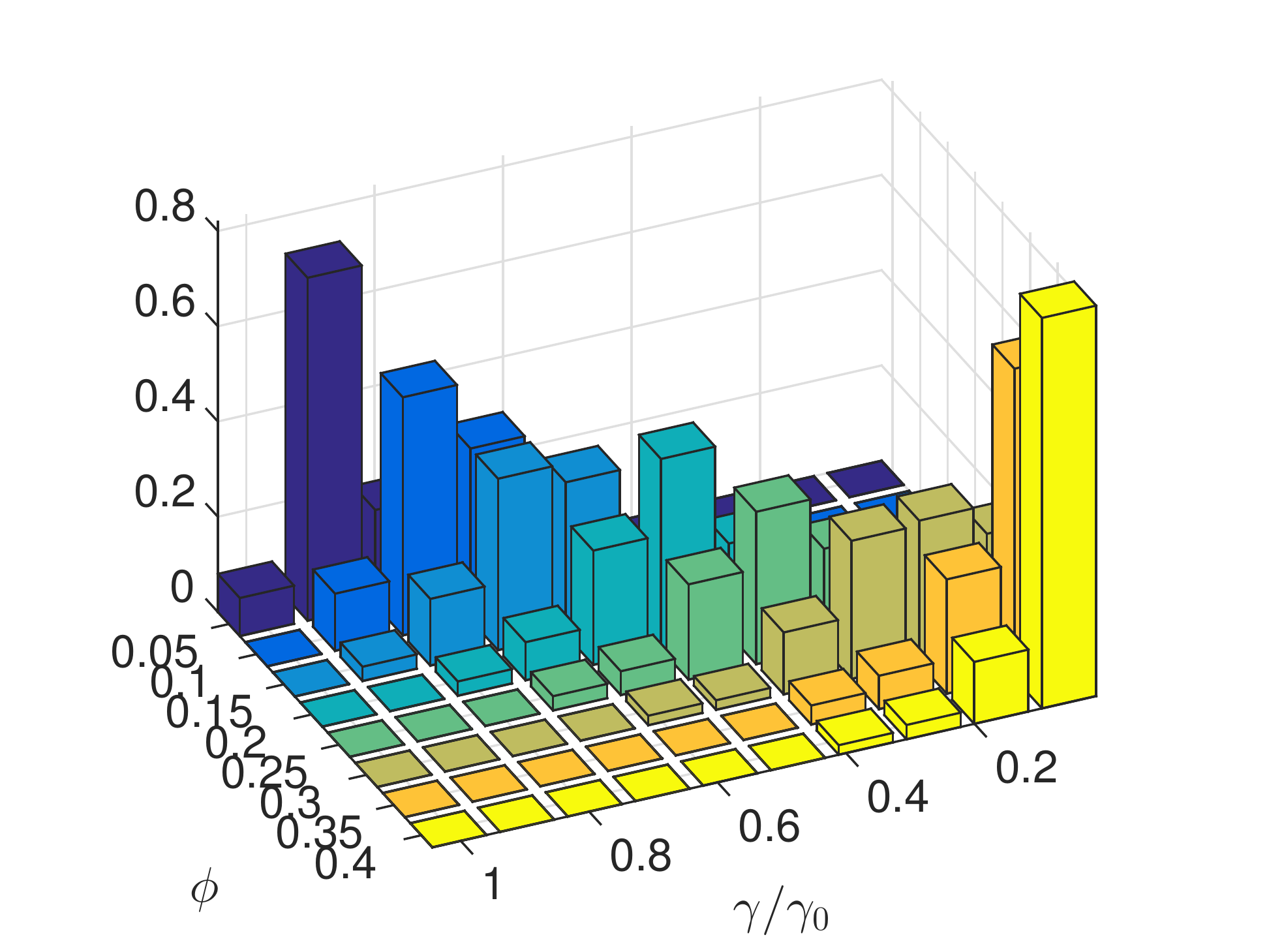}}
  \caption{Histogram counts of $\gamma/\gamma_0$ for different
    fractions of occupied volume $\phi$ based on $100$ different
    crowder distributions in 2D. The radii of the crowders and the
    tracer are $R/\rho = 0.1$ and $r/\rho = 0.1$. When $0.1 < \phi <
    0.25$ the $\gamma$-distributions are close to normal. When $\phi >
    0.3$ the molecule will not reach $\pomega_\ast$ implying that
    $E(\fatx)\rightarrow\infty$ and $\gamma/\gamma_0\rightarrow 0$ in
    \eqref{eq:gammaform} for many obstacle configurations.}
  \label{fig:gammadist}
\end{figure}

If the obstacles are mobile, it would be too expensive computationally
to determine a $\gamma_\alpha(t)$ in every time step of a discretized
equation \eqref{eq:diffequ1} and also all details of how the obstacles
are moving are likely not known. Instead, $\gamma$ is sampled from a
stationary distribution. This distribution is computed with a circle
or sphere of radius $\rho$ circumscribing a voxel of a typical size in
the mesh. The tracer and the obstacles are spheres with radii $r$ and
$R$, respectively. The obstacles are randomly distributed inside $\omega_\ast$
for a given percentage of occupied volume $\phi$. Then we compute
$\gamma$ by evaluating \eqref{eq:gammaform} at the center and collect
statistics. These distributions effectively approximate the PDF
$p_\gamma(\gamma|\phi)$, see Figure~\ref{fig:gammadist}.
The joint distribution for $\gamma$ and $\phi$ can be determined if
the PDF $p_\phi(\phi)$ of $\phi$ is known.  

The effects of a deterministic $\phi(\fatx)$ variable in space are
studied in \cite{SmCiGr17}.  Sampling a new $\gamma$ for the moving
molecule after $\Delta t$ accounts in \cite{SmCiGr17} for the movement
of the crowder molecules during that time step. This $\gamma$ sampling
corresponds to the molecules switching their internal states, and we
will couple the statistics in Figure~\ref{fig:gammadist} to $\fatA$
and $\fatT$ in \eqref{eq:diffeqy2} in the next section.

\subsection{Diffusion coefficients in internal states}
\label{sec:diffintst}

A molecule in different internal states $k$ in Section
\ref{sec:intstates} has different diffusion coefficients $\gamma_k$
and switches its state according to $\kappa_0\fatA$. We sample
these $\gamma_k$ from the stationary distributions in Section
\ref{sec:gamcomp} for a given $\phi$ and the frequency of the state
$k$ being $f_k$.

Let $\tau$ be the overall time scale for the speed of switching of the
internal states and let $\kappa_0=1/\tau$. A large $\tau$ with
$\kappa_0\ll\gamma_0$ implies that the time scale of switching the
internal states is slower than the scale of diffusion. A physical
interpretation is that the crowding obstacles move slowly and the
tracer hence diffuses with the same $\gamma_k$ for a long time. If
instead $\tau$ is small, then the motion of the obstacles is fast
compared to the tracer molecules.

The quotient between the diffusion coefficient with crowding
$\gamma_k$ in internal state $k$ obtained by coarse-graining and the
coefficient $\gamma_0$ in free space is denoted by
$\theta_k=\gamma_k/\gamma_0 \in [0,1]$. The diffusion in the $k$th
internal state in \eqref{eq:diffequ1}, \eqref{eq:diffequ2}, and
\eqref{eq:diffeqy1} is
\begin{equation}
  \gamma_k=\gamma_0\theta_k=\gamma_0 T_{kk}.
  \label{eq:diffintst}
\end{equation}
Hence, $T_{kk}=\theta_k$. Let the ordering of the internal states be
such that $\gamma_k<\gamma_{k+1}$.

The stationary distribution in the internal states is
$\mu_k/T_{kk}=\mu_k/\theta_k,\; k=1,\ldots,K,$ in \eqref{eq:uinf} and
\eqref{eq:Aprop}. We now set this stationary distribution proportional
to the frequency $f_k$ of the state $k$ computed by the homogenization
in Section~\ref{sec:gamcomp}
\begin{equation*}
\frac{\mu_k}{\theta_k}\propto f_k,
\end{equation*}
and after normalizing with $\sum_{j=1}^K\mu_j=1$ we obtain
\begin{equation}
  \mu_k=\frac{f_k\theta_k}{\sum_{j=1}^Kf_j\theta_j}.
\label{eq:mudef}
\end{equation}
The transfer matrix in \eqref{eq:Aspec} and \eqref{eq:Aprop} is defined by
\begin{equation}
  A_{ij}=\mu_i\theta_j,\; i\ne j,\quad A_{ii}=(\mu_i-1)\theta_i,
\label{eq:Adef}
\end{equation}
as in \cite{BEHL16}. The stationary probability $p_k=p(\gamma_k|\phi)$
to be in internal state $k$ is proportional to $f_k$ and
$\mu_k/\theta_k$. With a scaling such that $\sum_{j=1}^Kp_j=1$, we
have
\begin{equation}
  p_k=\frac{f_k}{\sum_{j=1}^Kf_j}=
  \frac{\mu_k/\theta_k}{\sum_{j=1}^K\mu_j/\theta_j}.
  \label{eq:pdef}
\end{equation}
Using \eqref{eq:diffintst} and \eqref{eq:pdef} the expected diffusion
rate for a molecule in the stationary state is
\begin{equation}
  \bgamma=\sum_{j=1}^K \gamma_j p_j=
  \gamma_0\frac{\sum_{j=1}^K f_j\theta_j}{\sum_{j=1}^K f_j}\le\gamma_0,
  \label{eq:statdiff}
\end{equation}
and the variance scaled by the square of the mean is 
\begin{equation}
  \frac{{\rm Var}[\gamma]}{\bgamma^2}=
  \bgamma^{-2}\sum_{j=1}^K (\gamma_j-\bgamma)^2 p_j=
  \sum_{j=1}^K\frac{\mu_j}{\theta_j}\sum_{j=1}^K \mu_j\theta_j-1.
\label{eq:varstatdiff}
\end{equation}
The mean diffusion coefficient $\bgamma$ in \eqref{eq:statdiff} is
reduced compared to diffusion in free space $\gamma_0$ if at least one
$\theta_k<1$. Both $\fatA$ in \eqref{eq:Adef} and $\bgamma$ in
\eqref{eq:statdiff} are determined uniquely by $\gamma_k$ and the
corresponding $f_k$.

The statistics in Figure~\ref{fig:gammadist} can be used to introduce
more internal states to represent also different crowding densities
$\phi_j$.  Both $\gamma_i$ and $\phi_j$ are then sampled from the
joint distribution by changing the internal states.

\begin{figure}[H]
  \center{\subfigure[]{\includegraphics[width=.55\textwidth]{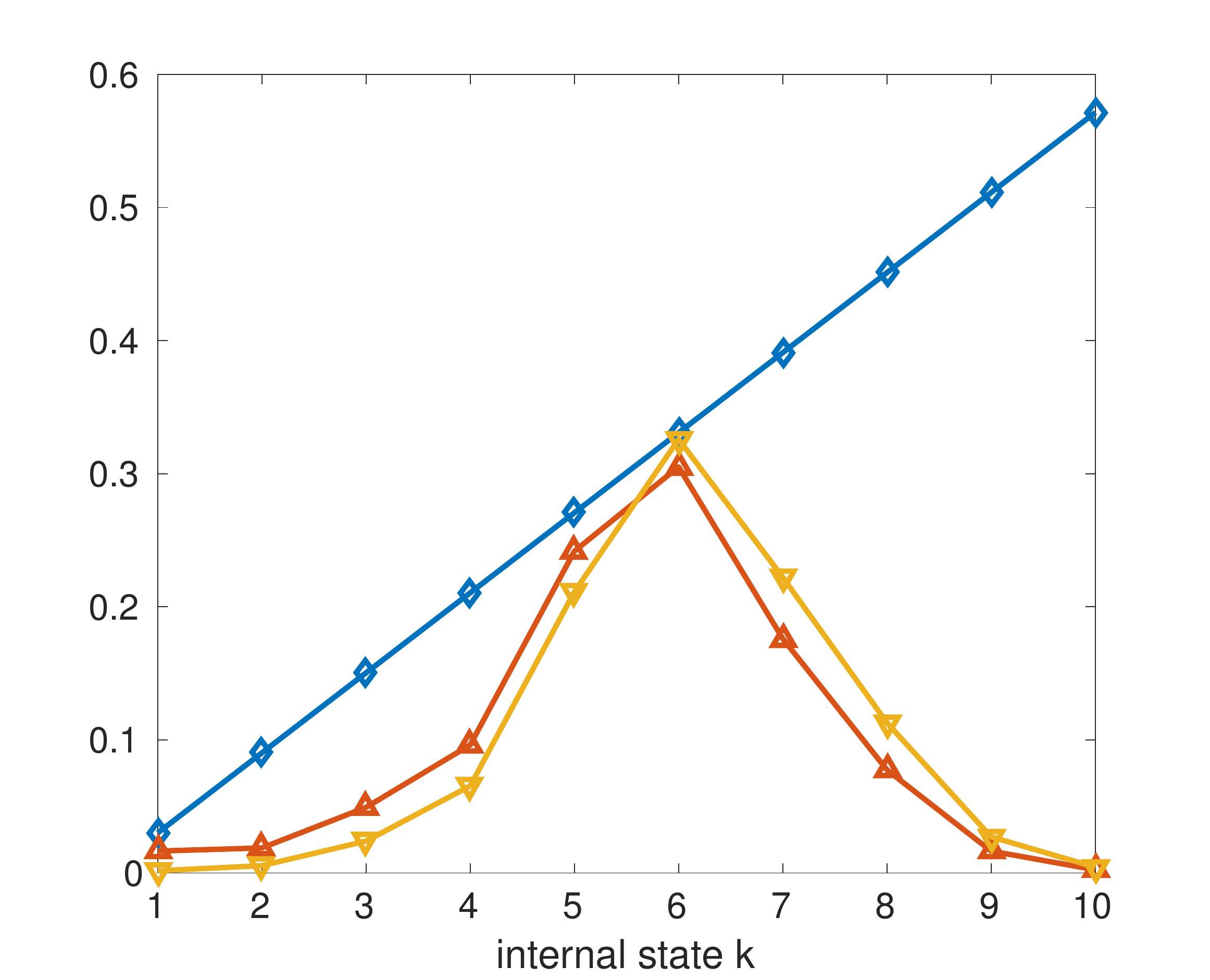} }}
  \caption{The computed $\theta_k\propto\gamma_k$ ($\diamond$, blue),
    $p_k$ ($\triangle$, red), and $\mu_k$ ($\nabla$, orange) for
    $k=1,\ldots,10,$ at $\phi=0.2$ in 2D.}
  \label{fig:mu}
\end{figure}

The frequencies $f_k$ for $\phi=0.2$ and $\gamma_k/\gamma_0$ determine
$\theta_k$, $p_k$, and $\mu_k$ in ten bins in an example in
Figure~\ref{fig:mu} using the statistics in
Figure~\ref{fig:gammadist}.  Since $\mu_k$ is proportional to $f_k$ in
\eqref{eq:mudef}, $\mu_k$ has the same support and is similar to $p_k$
in the bins.  With this $\fatp$, the distribution of $\gamma$ is well
approximated by a normal distribution $\calN(\bgamma, {\rm
  Var}[\gamma])$ with the mean and variance in \eqref{eq:statdiff} and
\eqref{eq:varstatdiff}.

\section{Analytical distributions}
\label{sec:analyt}

Consider an open chemical system with the monomolecular reactions
degradation, conversion, and production from a source and include
diffusion between voxels $i$ and $j$ and a switch of internal states
between $k$ and $l$. Then the transformations of the species are
\begin{equation}
    A_{ik}\mathop{\rightleftharpoons}A_{jk},\; A_{ik}\mathop{\rightleftharpoons}A_{il},\; A_{ik}\longrightarrow B_{ik},\; A_{ik}\longrightarrow\emptyset,\; 
    \emptyset\longrightarrow A_{ik}.
\label{eq:monoreact}
\end{equation}                                                         
In order from left to right the reactions in \eqref{eq:monoreact} are:
change of voxel by diffusion, change of internal state in a voxel,
conversion from $A$ to $B$ in the same voxel and internal state,
degradation of $A$, and production of $A$. When the reaction
propensities are independent of or linear in the copy numbers, the
expression for the probability distribution of molecules solving the
reaction-diffusion master equation is known explicitly at the
stochastic, mesoscopic level of modeling, see \cite{GaLeOt05, JaHu07}.
The analytical solutions of the PDFs of the chemical networks in
\eqref{eq:monoreact} are in this section used to derive the
macroscopic diffusion coefficient in \eqref{eq:hdiffcoeff} and the
statistical properties of the random molecular numbers of the species
in the steady state.

A random vector $\fatY$ with entries $Y_{ik\ell}$ is the state vector
for the copy number of species $\ell$ in internal state $k$ in voxel
$i$.  The mean value of $\fatY$ is denoted by $\fatby$ in Section
\ref{sec:discrdiff}. We determine the probability distribution of
$\fatY$ analytically for the transformations in \eqref{eq:monoreact}.

If the chemical system has the monomolecular reactions conversion and
degradation as in \eqref{eq:monoreact} except for the production of
$A$, then the system is closed and $\fatf$ in \eqref{eq:reacteq} in
voxel $i$ in internal state $k$ is
\begin{equation}
    \fatf(\fatby_{ik})=\fatR\fatby_{ik},
\label{eq:linfdef}
\end{equation}
where $\fatR$ is constant and $\fatR\in\mR^{L\times L}$.  For such an
$\fatf$ we have the stationary solution $\fatby_{ik\infty}$ satisfying
\begin{equation}
     \fatR\fatby_{ik\infty}=\fat0.
\label{eq:Ruinf}
\end{equation}
Initially, there are $N$ molecules in the system.  If $\emptyset$ is
regarded as a special species, then the number of molecules $N$ is
constant. No molecules are created and no molecules disappear.  If
there is degradation, the system may end up with all molecules in
$\emptyset$.

The PDF of the multinomial distribution $\calM(N, \fatp)$ for $\faty$
with $M=JKL$ states is
\begin{equation}
    p_{\calM}(\faty, \fatp)=\frac{N!}{y_1!\cdot y_2!\cdots y_M!}p_1^{y_1} p_2^{y_2} \cdots p_M^{y_M},\; \sum_{m=1}^M y_m=N.
\label{eq:multinompdf}
\end{equation}                                                         
Here $m$ is the global index $m=1+(j-1)+(k-1)J+(l-1)JK$ for the state
$jkl$ to simplify the notation.  The probability for a molecule to be
in state $m$ at $t$ is $p_m(t)$ and hence
$\sum_{m=1}^M p_m=1,\; p_m\ge 0.$ Assume that the initial distribution
of $\fatY$ at $t=0$ in the chemical system is $\calM(N,
\fatp_0)$.
Then it is proved in \cite{JaHu07} that the joint distribution of
$\fatY$ for all molecules for $t>0$ is multinomial
${\calM}(N, \fatp(t))$ where $\fatp(t)$ solves
\begin{equation}
    \frac{d\fatp}{dt}=\fatB \fatp,\quad \fatB=\gamma_0\fatLambda\Ktimes\fatT\Ktimes\fatI_L+\kappa_0\fatI_J\Ktimes\fatA\Ktimes\fatI_L
                                      +\fatI_J\Ktimes\fatG\Ktimes\fatR,
\label{eq:pMeq}
\end{equation}                                                         
with initial data $\fatp(0)=\fatp_0$. The system matrix $\fatB$ is
identical to the one in \eqref{eq:reacteq} where
\[
   \fate_J\Ktimes\fatg\Ktimes\fatf=\fatI_J\Ktimes\fatG\Ktimes\fatR\fatp
\]  
for our monomolecular reactions.

Let $\calW_w,\, w=1,\ldots, W,$ be subsets of
$\calI_M=\{1,2,\ldots,M\}$ such that $\bigcup_{w=1}^W\calW_w=\calI_M$
and introduce
\begin{equation}
    Z_w=\sum_{m\in\calW_w} Y_m,\quad q_w=\sum_{m\in\calW_w}p_m, \quad \fatz\in\mR^W.
\label{eq:Zqdef}
\end{equation}
Then by the properties of the multinomial distribution, the PDF of
$\fatZ$ is
\begin{equation}
    P(t, \fatz)=p_{\calM}(\fatz, \fatq(t)).
\label{eq:condmultinompdf}
\end{equation}                                                         
In particular, if $\fatz\in\mN^2$, i.e. $W=2$ and $z_w$ is integer and non-negative, then the distribution is binomial.

The stationary distribution when $t\rightarrow \infty$ is
\begin{equation}
   \lim_{t\rightarrow\infty}P(t,\faty)=p_{\calM}(\faty, \fatp_{\infty}),
\label{eq:statpdf}
\end{equation}                                                         
where $\fatp_{\infty}$ is the solution of  
\begin{equation}
    \fatB\fatp_\infty=0.
\label{eq:pinfdef}
\end{equation}
The vectors $\fatv$ and $\fatT^{-1}\fatmu$ satisfy $\fatLambda\fatv=\fat0$ and $\fatA\fatT^{-1}\fatmu=\fat0$ as in \eqref{eq:diffeqy2inf}.
Let $\fatp_{\Lambda\infty}, \fatp_{A\infty},$ and $\fatp_{R\infty}$ satisfy
\begin{equation}
\fatp_{\Lambda\infty}=\eta_\Lambda\fatv,\; \fatp_{A\infty}=\eta_A\fatT^{-1}\fatmu,\; \fatR\fatp_{R\infty}=\fat0,
\label{eq:pDARinf}
\end{equation}
with $\fatp_{R\infty}$ and scalings $\eta_\Lambda$ and $\eta_R$ chosen to fulfill
$\|\fatp_{\Lambda\infty}\|_1=\|\fatp_{A\infty}\|_1=\|\fatp_{R\infty}\|_1=1$.
The stationary distributions $\fatp_{\Lambda\infty}, \fatp_{A\infty},$
and $\fatp_{R\infty}$ are all independent of $\gamma_0$ and
$\kappa_0$.  If the reaction matrix $\fatR$ is irreducible such that
the chemical network cannot be decomposed into two or more independent
networks, then there is a $\fatp_{R\infty}$ with non-negative
components $p_{R\infty,i}$ in \eqref{eq:pDARinf} \cite{GaLeOt05,
  JaHu07}. It follows from \eqref{eq:pinfdef}, \eqref{eq:pMeq}, and
\eqref{eq:pDARinf} that
\begin{equation}
    \fatp_\infty=\fatp_{\Lambda\infty}\Ktimes\fatp_{A\infty}\Ktimes\fatp_{R\infty},\; \|\fatp_{\infty}\|_1=1.
\label{eq:pinf}
\end{equation}
 
With the conversion reaction in \eqref{eq:monoreact}, $\fatR$ is such
that $\fate_L^T\fatR=\fat0$. It follows from \eqref{eq:diffsumy} and
\eqref{eq:pMeq} that
\begin{equation}
    (\fate_J\Ktimes\fate_K\Ktimes\fate_L)^T\fatB=\fat0.
\label{eq:Bzero}
\end{equation}
Using \eqref{eq:pMeq}, we find that
\begin{equation}
    (\fate_J\Ktimes\fate_K\Ktimes\fate_L^T\fatp)_t=(\fate_J\Ktimes\fate_K\Ktimes\fate_L)^T\fatp_t=\fat0,
\label{eq:pderiv}
\end{equation}
and the probability is preserved
\begin{equation}
    \|\fatp(t)\|_1=\|\fatp(0)\|_1=\|\fatp_\infty\|_1=1,
\label{eq:pconst}
\end{equation}
with a properly chosen initial solution $\fatp_0=\fatp(0)$.

When the time scale of the diffusion is fast with a large
$\tau=\kappa_0^{-1}\gg \gamma_0^{-1}$, then by \eqref{eq:Adef}
$\kappa_0\fatA$ is negligible in \eqref{eq:pMeq} since $\fatA$ is of
$\calO(1)$.  Spatial gradients disappear rapidly and the system is
well-stirred.  On the contrary, if $\tau$ is small then
$\kappa_0\fatA$ dominates and there is a fast equilibration in the
internal states such that the solution is (after reordering the
unknowns $p_{ik\ell}$)
$\fatp(t)\approx\fatp'(t)\Ktimes\fatp_{A\infty}$ and
$\fatI_J\Ktimes\fatI_L\Ktimes\fatA\fatp\approx(\fatI_J\Ktimes\fatI_L)\fatp'\Ktimes\fatA\fatp_{A\infty}=\fat0$
in the second term in $\fatB$ in \eqref{eq:pMeq}.

The expected value $u_{ik\ell}$ of the concentration of species $\ell$
in voxel $i$ and internal state $k$ is given by
\begin{equation}
    u_{ik\ell}(t)=E\left[\frac{Y_{ik\ell}}{|\calV_i|}\right]=\frac{N}{|\calV_i|}p_{ik\ell}(t)=\frac{\by_{ikl}}{|\calV_i|}.
\label{eq:uexp}
\end{equation}
Since the mean values of the copy numbers $\fatoy(t)$ satisfy
\eqref{eq:pMeq}, $\fatu$ in \eqref{eq:uexp} satisfies an equation like
\eqref{eq:diffequ1} with an additional reaction term.

The diffusion coefficient in the equation for the observable
$U_{i\ell}$ in \eqref{eq:diffeqUi2} in voxel $i$ and species $\ell$
with $\fate^T_K\fatp_{i\ell}(t)>0$ is by \eqref{eq:hdiffcoeff},
\eqref{eq:uexp}, and \eqref{eq:diffintst}
\begin{equation}
    \hgamma_{i\ell}(t)=\gamma_0\frac{\fate^T_K\fatT\fatp_{i\ell}(t)}{\fate^T_K\fatp_{i\ell}(t)}
                    =\gamma_0\frac{\sum_{k=1}^K\theta_kp_{ik\ell}(t)}{\sum_{k=1}^K p_{ik\ell}(t)}\le \gamma_0,
\label{eq:hgam_u}
\end{equation}
since $0\le\theta_j\le 1$, cf. \eqref{eq:statdiff} for the stationary
case.  The time dependent diffusion coefficient $\hgamma_{i\ell}(t)$
is bounded from above by the nominal coefficient $\gamma_0$ and as
$t\rightarrow\infty$, $\hgamma_{i\ell}(t)$ approaches $\bgamma$ in
\eqref{eq:statdiff}.

A simpler alternative to $\hgamma$ in \eqref{eq:hgam_u} is to derive
the random diffusion field in \eqref{eq:diffeqUi2} as follows.  First,
discretize the time derivative in \eqref{eq:diffeqUi2} at
$t^n,\, n=0, 1,\ldots$, and sample $\hgamma_j^n$ with the stationary
distribution in \eqref{eq:pdef}. Then we have a numerical
approximation of the parabolic PDE \eqref{eq:Ueq2} discretized by
finite elements in \eqref{eq:diffeqUi2} with a random, space and time
dependent diffusion coefficient field $\tgamma$ with mean and variance
\eqref{eq:statdiff} and \eqref{eq:varstatdiff}.

The sum of the molecules over the internal states in each voxel and
for each species is denoted by
\begin{equation}
    Z_{i\ell}=\sum_{k=1}^K Y_{ik\ell}.
\label{eq:Zdef}
\end{equation}
Since $\fatY$ is multinomially distributed with parameters $\fatp$,
$\fatZ$ is also multinomially distributed $\calM(N, \fatq)$ according
to \eqref{eq:Zqdef} and \eqref{eq:condmultinompdf} where $\fatq$ has
the components
\begin{equation}
    q_{i\ell}(t)=\sum_{k=1}^K p_{ik\ell}(t).
\label{eq:qdef}
\end{equation}
At the stationary distribution, $\fatq$ is
\begin{equation}
    q_{i\ell\infty}=\sum_{k=1}^K p_{\Lambda\infty,i}p_{A\infty,k}p_{R\infty,\ell}=p_{\Lambda\infty,i}p_{R\infty,\ell}.
\label{eq:qinfdef}
\end{equation}
The observable $U_{i\ell}$ is the expected value of the concentration
of species $\ell$ in $\calV_i$
\begin{equation}
    U_{i\ell}(t)={\rm E}\left[\frac{Z_{i\ell}}{|\calV_i|}\right]=\frac{N}{|\calV_i|}q_{i\ell}(t).
\label{eq:Uexp}
\end{equation}
Using \eqref{eq:qinfdef}, we find that the steady state solution
$U_{\infty, i\ell}$ is independent of $i$ and thus constant in
space. The variance of the concentration is
\begin{equation}
    {\rm Var}\left[\frac{Z_{i\ell}}{|\calV_i|}\right]=\frac{N}{|\calV_i|^2}q_{i\ell}(t)(1-q_{i\ell}(t)).
\label{eq:VarU}
\end{equation}
The number of voxels $J$ is often large making $q_{i\ell}(t)\propto
1/JL$ and small and the variance is approximately
$Nq_{i\ell}(t)/|\calV_i|^2=U_{i\ell}/|\calV_i|$.  The covariance
between species $\ell$ in voxel $i$ and species $m$ in voxel $j$ is
\begin{equation}
    {\rm Cov}\left[\frac{Z_{i\ell}}{|\calV_i|}, \frac{Z_{jm}}{|\calV_j|}\right]=-\frac{N}{|\calV_i||\calV_j|}q_{i\ell}(t)q_{jm}(t).
\label{eq:CovU}
\end{equation}
The co-variation between the
voxels is negative and since $q_{i\ell}$ is usually small, it is very small. The mean and the variance of the copy numbers $Z_{i\ell}$ are
\begin{equation}
    E[Z_{i\ell}]=Nq_{i\ell}(t),\quad {\rm Var}[Z_{i\ell}]=Nq_{i\ell}(t)(1-q_{i\ell}(t)).
\label{eq:meanZ}
\end{equation}
The Fano factor ${\rm Var}[Z_{i\ell}]/E[Z_{i\ell}]$ is $1-q_{i\ell}(t)$ and close to 1, which is the factor of a Poisson process.
 
A similar analysis is possible for a chemical system when all
monomolecular reactions in \eqref{eq:monoreact} are included. If the
copy numbers $\fatY$ in the states of the system are Poisson
distributed initially then they will remain Poisson distributed with
rate parameters satisfying an equation like \eqref{eq:pMeq} and
\eqref{eq:reacteq}, see \cite{JaHu07}.



\section{Numerical examples}
\label{sec:examples}

We now proceed to illustrate the behavior of the suggested
coarse-grained model of subdiffusion in stochastic simulation of
trajectories of the chemical network. After first briefly summarizing
the simulation algorithm in Section~\ref{subsec:SSA}, we look at the
mean-square displacement of subdiffusing molecules on a circle in
Section~\ref{subsec:pure_subdiffusion} using a finite element
discretization over a triangular mesh to discretize the required
diffusion operator as in Section~\ref{sec:discrdiff}. In
Section~\ref{subsec:bimolecular}, we investigate the available range
of dynamics when bimolecular reactions are included. Finally, in
Section~\ref{subsec:mincde} we look at potential subdiffusive effects
when simulating a realistic three-dimensional model of a subsystem of
an \textit{E.~coli} model. In all examples, the mesoscopic internal
states model with variable diffusion coefficients is determined as in
Section~\ref{sec:Combine}.  With repeatability and reproducibility in
mind, the models tested here will be released in the coming version
1.4 of our freely available software URDME \cite{URDMEpaper,
  URDMEsoftware}.

\subsection{Stochastic Simulation Algorithm}
\label{subsec:SSA}

The direct simulation method \cite{gillespie} by Gillespie
determines the time for the next reaction event and which event that
will take place. For spatial problems the state of the chemical system
is a random variable $\fatY\in\mN^{JKL}$ and is defined by the number
of molecules of each species in the internal states in each voxel. The
simulation method of choice is then the next subvolume method (NSM)
\cite{Elf2004}. The probabilities for the events are given by the
coefficients in $\fatLambda\Ktimes\fatT$ (diffusion), $\fatA$ (change
of internal state), and the reaction propensities in $\fatf$. The
change of internal state in a voxel has the form of a monomolecular
reaction.

The NSM algorithm becomes time-consuming with multiple internal states
since many events simply change the internal states without advancing
the observable dynamics. A parallel version suitable for modern
multicore computers was developed in \cite{PNSM} which is 
effective in dealing with events taking place \emph{within} spatial
subdomains rather than between them. We remark that introducing the
internal states is a way of simulating a system with a random,
predetermined diffusion coefficient $\gamma(\fatx, t)$. Simulation of
such a system without internal states requires a special, more
complicated version of Gillespie's algorithm to handle time dependent
coefficients \cite{GibsonBruck}.

\subsection{Pure subdiffusion}
\label{subsec:pure_subdiffusion}

There is experimental evidence that the diffusive transport of
molecules in cells is sometimes anomalous \cite{HofFra, Krapf15,
  METZLER00}.  Let $\langle \cdot\rangle$ denote the average over the
trajectories of the molecules. The mean square displacement (MSD) of a
molecule at $\fatx(t)$ at time $t$ released at $\fatx(0)=0$ at $t=0$
behaves as
\begin{equation}
  \langle \|\fatx(t)\|^2_2\rangle\propto t^\alpha,
\label{eq:MSD}
\end{equation}
where $\alpha=1$ for ordinary diffusion and $\alpha \in (0,1)$ in
subdiffusion where, at least in a time interval shortly after $t=0$,
the molecules diffuse anomalously, see \cite{HofFra}. The reason for
the subdiffusion may be crowding effects by other molecules and the
process is then non-ergodic with a memory, see e.g.~\cite{HiAn95,
  MARQUEZLAGO12}.

The macroscopic observable $\fatU(\fatx, t)\in\mR^L$, e.g., the
concentrations of the chemical species, satisfies a diffusion equation
with a fractional time derivative \cite{METZLER00}
\begin{equation}
  \displaystyle{\frac{\partial \fatU}{\partial t}=\frac{\partial^{1-\alpha} }{\partial t^{1-\alpha}}(\gamma\Delta \fatU),}
\label{eq:FPDE}
\end{equation} 
at least in a time interval, $t\in[t_0, t_1]$. The fractional
derivative is defined according to Riemann-Liouville. The internal
state parameters $\fatmu$ and $\fattheta$ in \eqref{eq:Adef} are
determined by $\alpha$ in the FPDE in \cite{BEHL16, MOMMER09}.  Here
they are given by statistics obtained with the microscopic model in
\cite{Meinecke16}.

We compute the MSD \eqref{eq:MSD} of the internal states model by
coarse-graining into ten states as in Figure~\ref{fig:mu} following
the procedure described in Section~\ref{sec:diffintst} using $\phi =
0.2$ and $\gamma_0 = 0.01$.  The geometry is the unit circle and the
molecules are released at time $t = 0$ in the center and in the
fastest diffusing internal state with $\gamma_{10} =
\gamma_0\theta_{10}$. Since there are no reactions, all transition
rates act linearly and the moment equations are closed such that the
mean square displacement can be accurately determined by solving
\eqref{eq:pMeq} numerically for $\fatp$ with $L=1$ and $\fatR=1$ for
the probability to be in voxel $i$.  Then the MSD in \eqref{eq:MSD} is
$\langle \|\fatx(t)\|^2_2\rangle=\sum_i p_i(t)\|\fatx_i\|^2_2$ where
$\fatx_i$ is the center of voxel $i$. 

\begin{figure}[ht]
  \centering
  \includegraphics{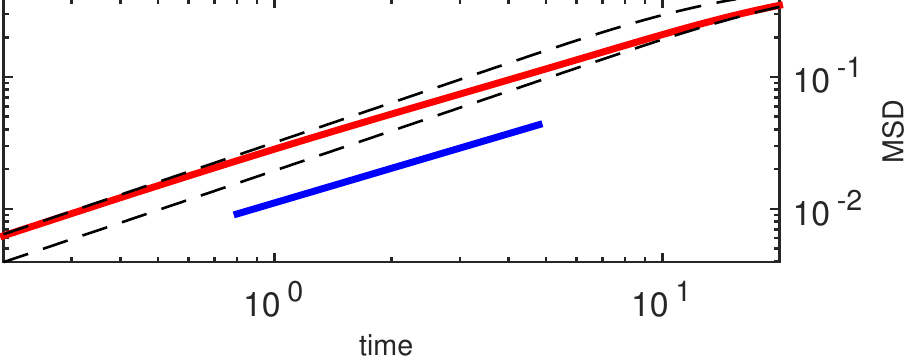}
  \includegraphics{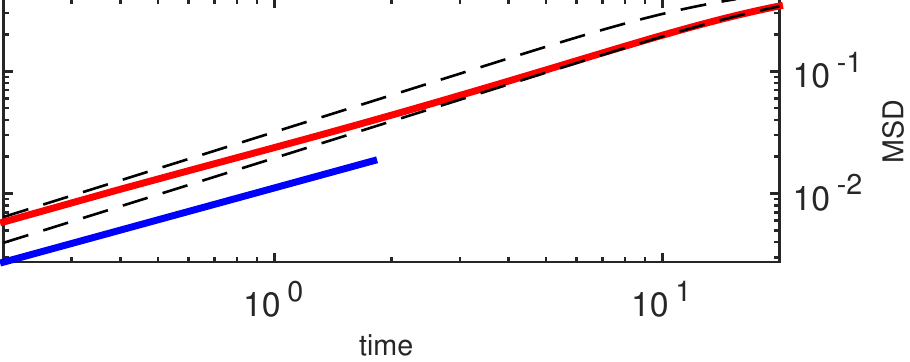}
  \caption{Simulation of coarse-grained subdiffusion in
    2D. \textit{Top:} The mean square displacement as a function of
    time (red). The dashed curves are obtained with the initial and
    the steady state diffusion, $\gamma_0$ and $\bgamma$, and the
    slope of the comparison curve $t^\alpha$ is $\alpha = 0.87$
    (blue).  \textit{Bottom:} As above, but with a four times faster
    diffusion of obstacles and hence faster scaling of time for the
    switching between internal states, $\kappa_0\rightarrow
    4\kappa_0$, resulting in a faster approach to the steady state
    diffusion (but still such that $\alpha = 0.87$ for the comparison
    slope). }
  \label{fig:pure2d_results}
\end{figure}

The initial diffusion rate is $\gamma_0\theta_{10}$ and, as $t \to
\infty$, $\gamma$ converges to $\bar{\gamma}$. The region in between
these two limits is where the subdiffusive behavior is observed, and
where $\alpha < 1$ in \eqref{eq:MSD} and \eqref{eq:FPDE}. Initially
and for large $t$, $\alpha=1$ and we have ordinary diffusion. By
scaling the internal transfer matrix $\fatA$ with a different
$\kappa_0$ in \eqref{eq:Aijkl}, this region (indicated by the
comparision slope) can be varied accordingly, see
Figure~\ref{fig:pure2d_results}, where the same $\alpha$ is obtained
with two different values of $\kappa_0$. Recall that $\kappa_0$ models
the speed of diffusion of the obstacles and for an accurate
description their average diffusion speed should be known.

\subsection{Bimolecular annihilation}
\label{subsec:bimolecular}

Consider two species $A$ and $B$ undergoing the single transition
\begin{align}
  A_i+B_j \to C,
\label{eq:ABCreact}
\end{align}
with $A$ in the internal state $i$ and $B$ in $j$ and with an
arbitrary internal state for $C$. Let the rate for this transition be
$H_{ij},\; i,j=1,\ldots,K$. Then the reaction propensity is
$H_{ij}a_ib_j$ where $a_i$ and $b_j$ are the copy numbers of $A_i$ and
$B_j$. Given an arbitrary non-negative rate matrix $\fatH$, a steady
state probability distribution $\fatp_{A\infty}$ of the internal
states for both $A$ and $B$, and a target rate constant $k_0$, we can
always scale $\fatH$ such that the mean rate agrees with the target at the steady state
\begin{equation}
    k_0=\fatp_{A\infty}^T\fatH\fatp_{A\infty}.
\label{eq:Ktarget}
\end{equation}
There is potentially great freedom in selecting the rate parameters
subject to a scaling. Note that when $H_{ij}$ is independent of $i$
and $j$ and $\fatH=k_0\fate_L\fate_L^T$, the internal states model
agrees with the standard model using a single target rate $k_0$. The
diffusion in the internal states is as in the previous example.

\begin{figure}[ht]
  \centering
  \includegraphics{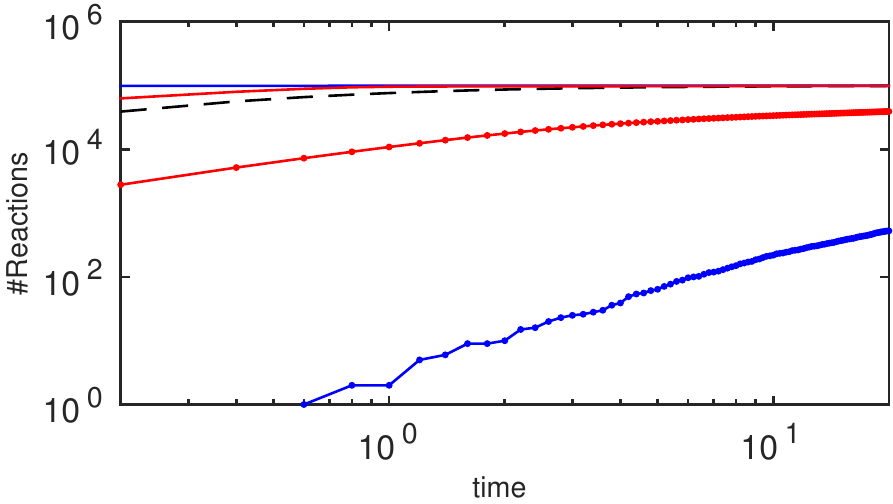}
  \caption{Results of the bimolecular reaction in \eqref{eq:ABCreact}
    presented as the time history of the number of resulting molecules
    $C$. The \emph{colored} lines represent
    simulations with $H_{ij}$ in the different cases 1 (blue with dots), 2 (red with dots), 3 (smooth red), and 4 (smooth blue),
    respectively. The \emph{dashed} line is the pure diffusion case
    with $k_0$ as the single rate.}
  \label{fig:annihilation_results}
\end{figure}

We release $A$ and $B$ molecules at time $t = 0$, $10^5$ of each
species, in ten internal states uniformly in space in the unit disc
with all in the fastest (the tenth) diffusing state. Four
different cases of rate parameters are defined as follows:
\begin{enumerate}
  \item $H_{1,1} = 1$ and 0 otherwise,
  \item $H_{ij} = (11-i)(11-j)$,
  \item $H_{ij} = ij$,
  \item $H_{10,10} = 1$ and 0 otherwise.
\end{enumerate}
Then the parameters are rescaled such that $\fatH$ satisfies
\eqref{eq:Ktarget} with $k_0 = 10^{-4}$. In cases 1 and 4, two
molecules $A$ and $B$ react only when they both are in the same voxel
and in the same internal state. The reaction rate decreases or
increases with the diffusion in cases 2 and 3. The combined effect of
internal states and reactions is modeled by $\fatH$ corresponding to
$\fatg\Ktimes\fatf$ in \eqref{eq:reacteq}.

The result obtained from a single realization of the system with
URDME, visualized as the number of resulting $C$ molecules, is
displayed in Figure~\ref{fig:annihilation_results}. The extreme cases 1 and 4 
where a single rate in $\fatH$ is non-zero are clearly identifiable,
as are the two intermediate cases 2 and 3. The single state model is found in
the middle of all of these cases. Different choices of reaction rates
yield a range of behavior. The idea that there is a freedom in
selecting the rate parameters opens up for advanced coarse-graining
methods based on, e.g., analytic and simulation results in the
diffusion-limited regime \cite{GrimaSchnell06, HLW06, Schnell2004}, or
computational methods based on data from Brownian dynamics
\cite{Andrews10, ZoWo5a} or molecular dynamics \cite{GROMACS}
simulations.

\subsection{\textit{Min} oscillations in \textit{E.~coli}}
\label{subsec:mincde}

\begin{figure}[ht]
  \centering
  \includegraphics{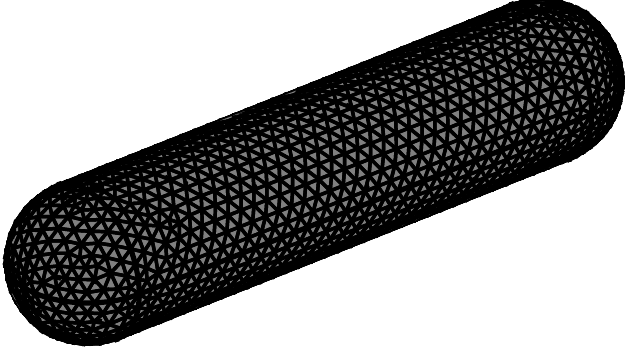}
  \caption{A discretized model of an \textit{E.~coli} bacterium. }
  \label{fig:ecoli}
\end{figure}

As a more involved example in three space dimensions, we take the
model from \cite{FaEl} of the \textit{Min}-system in the
\textit{E.~coli} bacterium. The geometry is rod-shaped with length
$3.5 \mu m$, diameter $1 \mu m$, and discretized using 9761
tetrahedra, see Figure~\ref{fig:ecoli}. MinD proteins oscillate from
pole to pole in the cell with a low concentration in the middle.
These oscillations help the cell locate its middle before cell
division \cite{Kruse:2002}. The five reactions, five species, and
reaction parameters from \cite{FaEl} are found in
Table~\ref{tab:MinD}. Two of the species, $\mbox{MinDmem}$ and
$\mbox{MinDE}$, are attached to the membrane and only diffuse
there. The other three species diffuse freely in the cytosol, where
the effective diffusion constant is $\gamma_0 = 2.5 \mu m^2/s$ in
\cite{FaEl}. Since the inside of an \textit{E.~coli} is a highly
crowded environment (cf.~Figure~\ref{fig:coli}), it is of interest to
investigate the incorporation of subdiffusion due to crowding and
reaction rates depending on the internal state in the mesoscopic
model.

\begin{table}[htp]
\centering
\begin{tabular}{ll}
  $\mbox{MinDcytATP} \xrightarrow{k_d} \mbox{MinDmem}$ & 
  $\mbox{MinDcytATP}+\mbox{MinDmem} \xrightarrow{k_{dD}} \mbox{2MinDmem}$ \\
  $\mbox{MinE+MinDmem} \xrightarrow{k_{de}} \mbox{MinDE}$ & 
  $\mbox{MinDE} \xrightarrow{k_e} \mbox{MinDcytADP}+\mbox{MinE}$ \\
  $\mbox{MinDcytADP} \xrightarrow{k_p} \mbox{MinDcytATP}$
 \end{tabular} 
 \caption{The chemical reactions of the {\it Min}-system. The
   constants take the values $k_d = 0.0125\mu m^{-1}s^{-1}$, $k_{dD} =
   9 \times 10^6 M^{-1}s^{-1}$ (here scaled by an additional factor of
   1.65 in the numerical experiments), $k_{de} = 5.56\times 10^7
   M^{-1}s^{-1}$, $k_e=0.7s^{-1}$, and $k_p = 0.5s^{-1}$.}
  \label{tab:MinD}
\end{table}

\begin{figure}
  \centering
  \includegraphics[width = 0.45\textwidth]{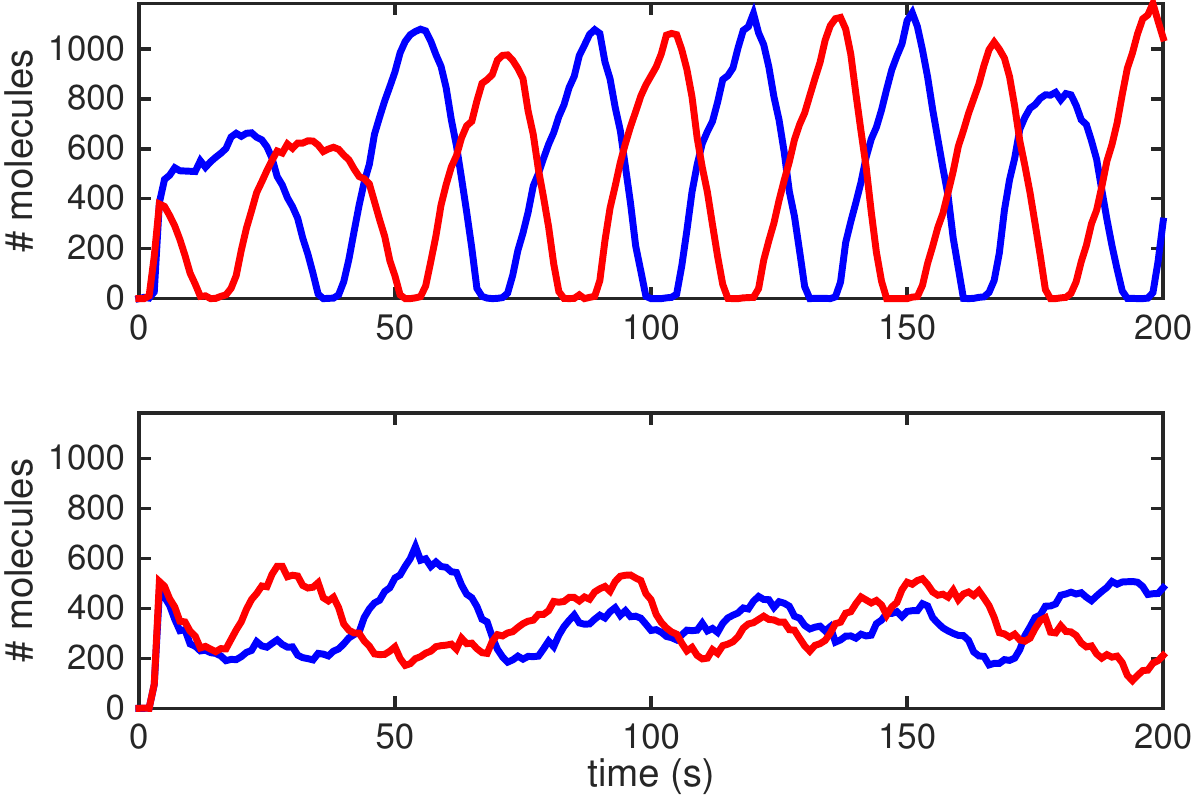}
  \includegraphics[width = 0.45\textwidth]{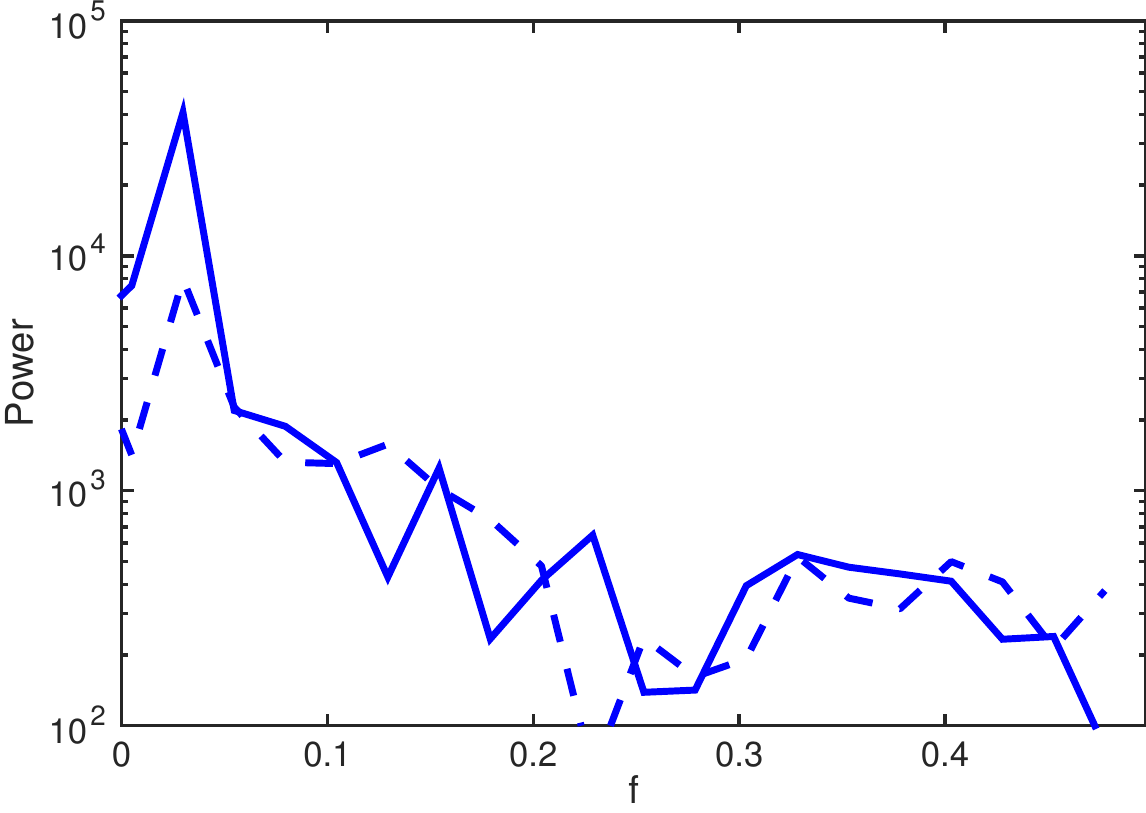}
  \caption{Two realizations of \textit{MinD}-oscillations in the
    membrane of an \textit{E.~coli} bacterium. \textit{Left:} The
    number of MinD molecules in the leftmost (red) and rightmost
    (blue) quarters of the bacterium, respectively.  \textit{Top:}
    Ordinary diffusion without internal states. \textit{Bottom:} Our
    coarse-grained subdiffusion model. \textit{Right:} The Fourier
    power spectrum of the pole oscillations of the two
    models : ordinary diffusion (solid) and 
    coarse-grained subdiffusion (dashed). }
  \label{fig:oscillations}
\end{figure}

As a proof-of-concept and in order to demonstrate the possibilities
here, we scaled the critical binding reaction rate $k_{dD}$ by a
factor 1.65, thus bringing the kinetics into a more sensitive regime
compared to the original rate. The normally diffusing model then
displays stable oscillations of the $Min$-protein in the membrane (see
upper left panel in Figure~\ref{fig:oscillations}).

As in the previous experiments we employ ten internal states obtained
from coarse-graining at $\phi = 0.2$. For the binding reaction of
state $i$, we multiply $k_{dD}$ by a factor $1+0.03i$ meaning that the
reactivity increases with faster diffusion. We then rescale the
resulting rate as in \eqref{eq:Ktarget} such that the steady state
mean rate agrees with the single state model. To bring in a bias we
arbitrarily let all reactions produce products in the fastest
diffusing state with $i=10$ ($A_i+B_j\to C_{10},\,\forall i,j$ where
$1\leq i,j\leq 10$), thus skewing the distribution over the internal
states towards faster diffusion and also faster binding rate. The
presence of subdiffusion and variable reaction rates in this model has
a striking effect on the oscillatory behavior. The oscillations are
damped considerably, see the lower left panel in
Figure~\ref{fig:oscillations}. The peak in the power spectrum at about
$0.03 Hz$ in the right panel of Figure~\ref{fig:oscillations} is
reduced by more than a factor 6 in the internal state model.  With our
computational framework, an investigation of the dynamics due to
crowding and variable reaction rates is computationally feasible even
in non-trivial and quite large examples.

\section{Conclusions}
\label{sec:conclusions}

\begin{figure}[H]
  \centering
  \includegraphics[width = 0.95\textwidth,clip = true,trim = 0 9cm 0
    1cm]{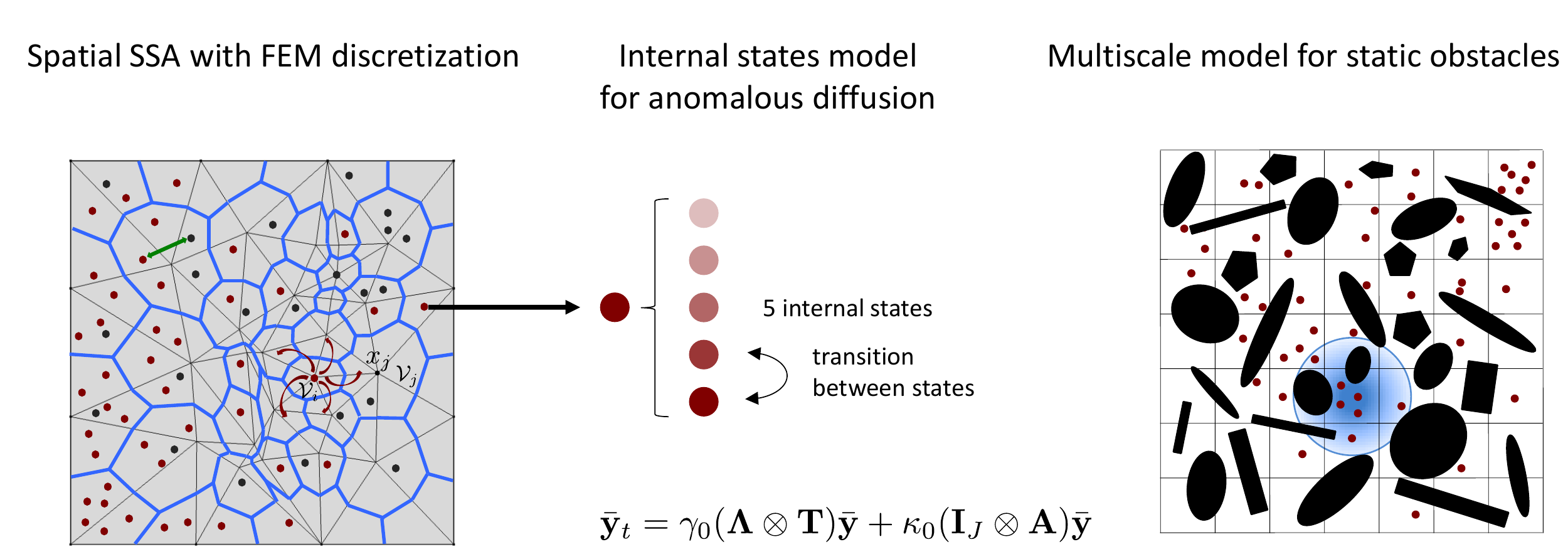}
  \caption{Summary of the method. \textit{Left:} A triangulation (grey edges) between the nodes $\fatx_j$ defines the primal mesh and the dual mesh (blue boundaries) which forms the voxels $\calV_j$.   Red and black tracer
    molecules diffuse (red arrows) between voxels. They react with each other (green arrow) when they are located in the same voxel.
    \textit{Middle:} The molecules
    in a voxel are in $K$ internal states ($K=5$ here). In the
    equation for the mean values $\fatby$ of the number of molecules
    in each state in every voxel, $\gamma_0$ is the free diffusion
    coefficient, $\fatLambda$ is the matrix of jump coefficients
    between the voxels given by a finite element discretization,
    $\fatT$ scales the hindered diffusion due to crowding in the $K$ internal states, $1/\kappa_0$ determines the time scale of the internal
    jumps, and $\fatA$ is the matrix of jump coefficients between the
    internal states in a voxel. \textit{Right:} The jump rates between
    the internal states in $\fatA$ are determined by computing the
    mean first exit time from the blue circle for a red tracer
    molecule with black obstacle molecules. Statistics is collected
    for many different obstacle configurations. Sampling from this distribution means that the tracer molecule is experiencing different crowder densities and consequently changes its internal state.
    \textit{Summary:} The novelty of our approach lies in coupling the internal states model (middle) which has previously been used to simulate anomalous diffusion to the explicit description of crowder molecules by coarse-graining the microscopic information to the mesoscopic level (right).}
\label{fig:summary}
\end{figure}

We have developed a computationally efficient approach to simulate diffusive and
subdiffusive transport processes on the mesoscopic level taking the explicit description of obstacle sizes and densities into account. 
We therefore couple two existing methods: the internal states model and the coarse-graining of a microscopic crowded geometry to the mesoscopic level, see the summary in Figure~\ref{fig:summary}.
Our novel method is faster than directly simulating microscopic Brownian dynamics and permits more detailed
modeling than a standard mesoscopic model with fixed diffusion and
reaction coefficients.  
In other lattice methods for simulation of
crowding, only a limited number of molecules can occupy the same voxel
in the lattice.  Compared to those methods, our method is less
heuristic and models the effect of crowding by deriving a distribution
of diffusion coefficients from a fine-grain geometry with obstacles of
different shape and size. 

An observable is the sum of the copy numbers in all internal states.
The mean values of the copy numbers of the observables satisfy
macroscopic PDEs discretized by a finite element method.  The
diffusion in the PDE for the observables is not explicitly known
unless the mean values of the full mesoscopic system are known.  

The crowding model has been implemented in URDME \cite{URDMEpaper,
  URDMEsoftware} and examples in 2D and 3D show the effects of
crowding and the modeling of the reactions.  The mean square
displacement of a diffusing molecule is computed and the $\alpha$
parameter measuring the deviation from Brownian motion is recorded.
Subdiffusive behavior is observed in a time interval after release of
the molecule.  The reaction propensities vary with the internal state
in two examples.  The scaling of the reaction coefficients is such
that the same steady state is reached but the transient phase differs
in the simulations depending on the particular choice of internal
representation.  This is illustrated in one example.  In the other
example, a realization of the {\it MinD} system without internal
states is oscillatory but is irregular with an internal structure in
the voxels.
  
The data for calibration of the internal states are here taken from
homogenization of a detailed microscopic model of crowding but other
sources are also possible. 
One alternative would be to infer the diffusion and reaction rates from  the posterior distribution of a Bayesian approach to analysis of experimental data. 
Another possibility would be to obtain the rates from coarse-graining data from Brownian
dynamics or molecular dynamics realizations of diffusion and
reactions. 

Microscopic and detailed computational methods are very
expensive for simulation of biochemical networks and are restricted to smaller subsystems and for short time. 
Our mesoscopic method including microscopic data offers a fast and
accurate approach for larger systems and longer time intervals at a
much reduced computational cost.

\section*{Acknowledgment}

The development of the software URDME (\url{www.urdme.org}) was
partially supported by the Swedish Research Council within the UPMARC
Linnaeus center of Excellence (S.~Engblom). Figure~\ref{fig:coli} was
kindly provided by Professor David van der Spoel, Uppsala University.

S.~Engblom and P.~L{\"o}tstedt would like to thank the Isaac Newton
Institute for Mathematical Sciences, Cambridge, for support and
hospitality during the programme Stochastic Dynamical Systems in
Biology: Numerical Methods and Applications where work on this paper
was undertaken. This work was supported by EPSRC grant no
EP/K032208/1.


\newcommand{\doi}[1]{\href{http://dx.doi.org/#1}{doi:#1}}
\newcommand{\available}[1]{Available at \url{#1}}
\newcommand{\availablet}[2]{Available at \href{#1}{#2}}


\begin{thebibliography}{10}

\bibitem{Luby-Phelps00}
K.~Luby-Phelps.
\newblock Cytoarchitecture and physical properties of cytoplasm: volume,
  viscosity, diffusion, intracellular surface area.
\newblock {\em Int. Rev. Cytology}, 192:189--221, 1999.

\bibitem{Schnell2004}
S.~Schnell and T.~E. Turner.
\newblock {Reaction kinetics in intracellular environments with macromolecular
  crowding: Simulations and rate laws}.
\newblock {\em Prog. Biophys. Mol. Biol.}, 85(2-3):235--260, 2004.

\bibitem{WoMu14}
P.~R. ten Wolde and A.~Mugler.
\newblock Importance of crowding in signaling, genetic, and metabolic networks.
\newblock {\em Int. Rev. Cell Mol. Bio.}, 307:419--442, 2014.

\bibitem{Grasberger86}
B.~Grasberger, A.~P. Minton, C.~DeLisi, and H.~Metzger.
\newblock {Interaction between proteins localized in membranes.}
\newblock {\em Proc. Natl. Acad. Sci. USA}, 83(17):6258--6262, 1986.

\bibitem{Jin2007}
S.~Jin and A.~S. Verkman.
\newblock {Single particle tracking of complex diffusion in membranes:
  Simulation and detection of barrier, raft, and interaction phenomena}.
\newblock {\em J. Phys. Chem. B}, 111(14):3625--3632, 2007.

\bibitem{Krapf15}
D.~Krapf.
\newblock Mechanisms underlying anomalous diffusion in the membrane.
\newblock {\em Curr. Topics Membr.}, 75:167--207, 2015.

\bibitem{Medalia2002}
O.~Medalia, I.~Weber, A.~S. Frangakis, D.~Nicastro, G.~Gerisch, and
  W.~Baumeister.
\newblock {Macromolecular architecture in eukaryotic cells visualized by
  cryoelectron tomography.}
\newblock {\em Science}, 298(5596):1209--1213, 2002.

\bibitem{DiRienzo2014}
C.~{Di Rienzo}, V.~Piazza, E.~Gratton, F.~Beltram, and F.~Cardarelli.
\newblock {Probing short-range protein Brownian motion in the cytoplasm of
  living cells}.
\newblock {\em Nat. Commun.}, 5:5891, 2014.

\bibitem{Galanti14}
M.~Galanti, D.~Fanelli, A.~Maritan, and F.~Piazza.
\newblock {Diffusion of tagged particles in a crowded medium}.
\newblock {\em Europhys. Lett.}, 107(2):20006, 2014.

\bibitem{HofFra}
F.~H{\"o}fling and T.~Franosch.
\newblock Anomalous transport in the crowded world of biological cells.
\newblock {\em Rep. Progr. Phys.}, 76:046602, 2013.

\bibitem{NiHaBu07}
{\noopsort{Nicolau}}{D. V. Nicolau Jr, J. F. Hancock, and K. Burrage}.
\newblock Sources of anomalous diffusion on cell membranes: {A} {M}onte {C}arlo
  study.
\newblock {\em Biophys. J.}, 92:1975--1987, 2007.

\bibitem{Hall2003}
D.~Hall and A.~P. Minton.
\newblock {Macromolecular crowding: Qualitative and semiquantitative successes,
  quantitative challenges}.
\newblock {\em Biochim. Biophys. Acta - Proteins Proteomics}, 1649(2):127--139,
  2003.

\bibitem{Ando10}
T.~Ando and J.~Skolnick.
\newblock {Crowding and hydrodynamic interactions likely dominate in vivo
  macromolecular motion.}
\newblock {\em Proc. Natl. Acad. Sci. USA}, 107(43):18457--18462, 2010.

\bibitem{Penington2011}
C.~J. Penington, B.~D. Hughes, and K.~A. Landman.
\newblock {Building macroscale models from microscale probabilistic models: A
  general probabilistic approach for nonlinear diffusion and multispecies
  phenomena}.
\newblock {\em Phys. Rev. E}, 84(4):041120, 2011.

\bibitem{Saxton07}
M.~J. Saxton.
\newblock A biological interpretation of transient anomalous subdiffusion. {I}.
  {Q}ualitative model.
\newblock {\em Biophys. J.}, 92:1178--1191, 2007.

\bibitem{YAL04}
S.~B. Yuste, L.~Acedo, and K.~Lindenberg.
\newblock Reaction front in an {$A+B\rightarrow C$} reaction-subdiffusion
  process.
\newblock {\em Phys. Rev. E}, 69:036126, 2004.

\bibitem{KiNgWaBoSuTa14}
D.~J. Kiviet, P.~Nghe, N.~Walker, S.~Boulineau, V.~Sunderlikova, and S.~J.
  Tans.
\newblock Stochasticity of metabolism and growth at the single-cell level.
\newblock {\em Nature}, 514:376--379, 2014.

\bibitem{McAArk97}
H.~H. McAdams and A.~Arkin.
\newblock Stochastic mechanisms in gene expression.
\newblock {\em Proc.~Natl.~Acad.~Sci.~USA}, 94(3):814--819, Feb. 1997.

\bibitem{PedOud05}
J.~M. Pedraza and A.~van Oudenaarden.
\newblock Noise propagation in gene networks.
\newblock {\em Science}, 307(5717):1965--1969, Mar. 2005.

\bibitem{ShaSwa08}
V.~Shahrezaei and P.~S. Swain.
\newblock The stochastic nature of biochemical networks.
\newblock {\em Curr. Op. Biotech.}, 19:369--374, 2008.

\bibitem{SwaElo02}
P.~S. Swain, M.~B. Elowitz, and E.~D Siggia.
\newblock Intrinsic and extrinsic contributions to stochasticity in gene
  expression.
\newblock {\em Proc.~Natl.~Acad.~Sci.~USA}, 99(20):12795--12800, Oct. 2002.

\bibitem{VanKampen}
{\noopsort{Kampen}}{N.~G.~van Kampen}.
\newblock {\em Stochastic Processes in Physics and Chemistry}.
\newblock Elsevier, Amsterdam, 2nd edition, 2004.

\bibitem{gillespie}
D.~T. Gillespie.
\newblock A general method for numerically simulating the stochastic time
  evolution of coupled chemical reactions.
\newblock {\em J.~Comput.~Phys.}, 22(4):403--434, 1976.

\bibitem{Elf2004}
J.~Elf and M.~Ehrenberg.
\newblock {Spontaneous separation of bi-stable biochemical systems into spatial
  domains of opposite phases}.
\newblock {\em Syst. Biol. IEE Proc.}, 1(2):230--236, 2004.

\bibitem{mesoRD}
J~Hattne, D~Fange, and J~Elf.
\newblock Stochastic reaction-diffusion simulation with {MesoRD}.
\newblock {\em Bioinformatics}, 21(12):2923--2924, 2005.

\bibitem{IsP}
S.~A. Isaacson and C.~S. Peskin.
\newblock Incorporating diffusion in complex geometries into stochastic
  chemical kinetics simulations.
\newblock {\em SIAM J.~Sci.~Comput.}, 28:47--74, 2006.

\bibitem{URDMEpaper}
B.~Drawert, S.~Engblom, and A.~Hellander.
\newblock {URDME}: a modular framework for stochastic simulation of
  reaction-transport processes in complex geometries.
\newblock {\em BMC Syst.~Biol.}, 6(76):1--17, 2012.

\bibitem{EnFeHeLo}
S.~Engblom, L.~Ferm, A.~Hellander, and P.~L\"{o}tstedt.
\newblock Simulation of stochastic reaction-diffusion processes on unstructured
  meshes.
\newblock {\em SIAM J.~Sci.~Comput.}, 31:1774--1797, 2009.

\bibitem{Burr17}
K.~Burrage, P.~Burrage, A.~Leier, and T.~Marquez-Lago.
\newblock A review of stochastic and delay simulation approaches in both time
  and space in computational cell biology.
\newblock In D.~Holcman, editor, {\em Stochastic Processes, Multiscale
  Modeling, and Numerical Methods for Computational Cellular Biology}, pages
  241--261, Cham, 2017. Springer.

\bibitem{EngHelLot17}
S.~Engblom, A.~Hellander, and P.~L{\"o}tstedt.
\newblock Multiscale simulation of stochastic reacion-diffusion networks.
\newblock In D.~Holcman, editor, {\em Stochastic Processes, Multiscale
  Modeling, and Numerical Methods for Computational Cellular Biology}, pages
  55--79, Cham, 2017. Springer.

\bibitem{MahmutovicElf}
A.~Mahmutovic, D.~Fange, O.~G. Berg, and J.~Elf.
\newblock Lost in presumption: stochastic reactions in spatial models.
\newblock {\em Nat. Meth.}, 9(12):1--4, 2012.

\bibitem{Roberts2013}
E.~Roberts, J.~E. Stone, and Z.~Luthey-Schulten.
\newblock {Lattice microbes: High-performance stochastic simulation method for
  the reaction-diffusion master equation}.
\newblock {\em J. Comput. Chem.}, 34(3):245--255, 2013.

\bibitem{Fanelli10}
D.~Fanelli and A.~J. McKane.
\newblock {Diffusion in a crowded environment}.
\newblock {\em Phys. Rev. E}, 82(2):021113, 2010.

\bibitem{Landman11}
K.~A. Landman and A.~E. Fernando.
\newblock {Myopic random walkers and exclusion processes: Single and
  multispecies}.
\newblock {\em Phys. A, Stat. Mech. Appl.}, 390(21-22):3742--3753, 2011.

\bibitem{Taylor2015}
P.~R. Taylor, C.~A. Yates, M.~J. Simpson, and R.~E. Baker.
\newblock {Reconciling transport models across scales: The role of volume
  exclusion}.
\newblock {\em Phys. Rev. E}, 92(4):040701, 2015.

\bibitem{Meinecke16}
L.~Meinecke.
\newblock Multiscale modeling of diffusion in a crowded environment.
\newblock {\em Bull. Math. Biol.}, 79:2672--2695, 2017.

\bibitem{Andrews10}
S.~S. Andrews, N.~J. Addy, R.~Brent, and A.~P. Arkin.
\newblock {Detailed simulations of cell biology with Smoldyn 2.1}.
\newblock {\em PLoS Comput. Biol.}, 6(3):e1000705, 2010.

\bibitem{ZoWo5a}
{\noopsort{Zon}}{J.~S.~van~Zon and P.~R. ten Wolde}.
\newblock Green's-function reaction dynamics: {A} particle-based approach for
  simulating biochemical networks in time and space.
\newblock {\em J.~Chem.~Phys.}, 123:234910, 2005.

\bibitem{SmithGrima17}
S.~Smith and R.~Grima.
\newblock Fast simulation of {B}rownian dynamics in a crowded environment.
\newblock {\em J. Chem. Phys.}, 146:024105, 2017.

\bibitem{MARQUEZLAGO12}
T.~T. Marquez-Lago, A.~Leier, and K.~Burrage.
\newblock Anomalous diffusion and multifractional {B}rownian motion: simulating
  molecular crowding and physical obstacles in systems biology.
\newblock {\em IET Syst. Biol.}, 6(4):134--142, 2012.

\bibitem{MeEr16}
L.~Meinecke and M.~Eriksson.
\newblock Excluded volume effects in on- and off-lattice reaction-diffusion
  models.
\newblock {\em IET Syst. Biol.}, 11(2):55--64, 2017.

\bibitem{BaGaMe12}
E.~Barkai, Y.~Garini, and R.~Metzler.
\newblock Strange kinetics of single molecules in living cells.
\newblock {\em Physics Today}, 65(8):29--35, 2012.

\bibitem{METZLER00}
R.~Metzler and J.~Klafter.
\newblock The random walk's guide to anomalous diffusion: a fractional dynamics
  approach.
\newblock {\em Phys. Rep.}, 339(1):1--77, 2000.

\bibitem{MOMMER09}
M.~S. Mommer and D.~Lebiedz.
\newblock Modeling subdiffusion using reaction diffusion systems.
\newblock {\em SIAM J. Appl. Math.}, 70(1):112--132, 2009.

\bibitem{BEHL16}
E.~Blanc, S.~Engblom, A.~Hellander, and P.~L{\"o}tstedt.
\newblock Mesoscopic modeling of stochastic reaction-diffusion kinetics in the
  subdiffusive regime.
\newblock {\em Multiscale Model. Simul.}, 14:668--707, 2016.

\bibitem{PLUE13}
F.~Persson, M.~Lind{\'e}n, C.~Unoson, and J.~Elf.
\newblock Extracting intracellular diffusive states and transition rates from
  single molecule tracking data.
\newblock {\em Nat. Meth.}, 10:265--269, 2013.

\bibitem{Oksendal}
B.~{\O}ksendal.
\newblock {\em Stochastic Differential Equations}.
\newblock Springer, Berlin, 6th edition, 2003.

\bibitem{GuWeZh14}
M.~D. Gunzburger, C.~G. Webster, and G.~Zhang.
\newblock Stochastic finite element methods for partial differential equations
  with random input data.
\newblock {\em Acta Numerica}, pages 521--650, 2014.

\bibitem{LoMe}
P.~L\"{o}tstedt and L.~Meinecke.
\newblock Simulation of stochastic diffusion via first exit times.
\newblock {\em J.~Comput.~Phys.}, 300:862--886, 2015.

\bibitem{MEHL15}
L.~Meinecke, S.~Engblom, A.~Hellander, and P.~L\"{o}tstedt.
\newblock Analysis and design of jump coefficients in discrete stochastic
  diffusion models.
\newblock {\em SIAM J.~Sci.~Comput.}, 38:A55--A83, 2016.

\bibitem{GaLeOt05}
C.~Gadgil, C.~H. Lee, and H.~G. Othmer.
\newblock A stochastic analysis of first-order reaction networks.
\newblock {\em Bull. Math. Biol.}, 67:901--946, 2005.

\bibitem{SmCiGr17}
S.~Smith, C.~Cianci, and R.~Grima.
\newblock Macromolecular crowding directs the motion of small molecules inside
  cells.
\newblock {\em J. R. Soc. Interface}, 14:20170047, 2017.

\bibitem{JaHu07}
T.~Jahnke and W.~Huisinga.
\newblock Solving the chemical master equation for monomolecular reaction
  systems analytically.
\newblock {\em J. Math. Biol.}, 54:1--26, 2007.

\bibitem{URDMEsoftware}
S.~Engblom et~al.
\newblock {URDME}: {U}nstructured {R}eaction-{D}iffusion {M}aster {E}quation,
  2008--2017.
\newblock {\it Multiple versions exist.} \available{www.urdme.org}.

\bibitem{PNSM}
J.~Lind{\'e}n, P.~Bauer, S.~Engblom, and B.~Jonsson.
\newblock Exposing inter-process information for efficient parallel discrete
  event simulation of spatial stochastic systems.
\newblock {\it ACM SIGSIM-PADS}, Singapore, May 24--26, 2017.

\bibitem{GibsonBruck}
M.~A. Gibson and J.~Bruck.
\newblock Efficient exact stochastic simulation of chemical systems with many
  species and many channels.
\newblock {\em J.~Phys.~Chem.}, 104(9):1876--1889, 2000.

\bibitem{HiAn95}
R.~Hilfer and L.~Anton.
\newblock Fractional master equations and fractal time random walks.
\newblock {\em Phys. Rev. E}, 51(2):R848--R851, 1995.

\bibitem{GrimaSchnell06}
R.~Grima and S.~Schnell.
\newblock A systematic investigation of the rate laws valid in intracellular
  environments.
\newblock {\em Biophys. Chem.}, 124:1--10, 2006.

\bibitem{HLW06}
B.~J. Henry, T.~A.~M. Langlands, and S.~L. Wearne.
\newblock Anomalous diffusion with linear reaction dynamics: From continuous
  time random walks to fractional reaction-diffusion equations.
\newblock {\em Phys. Rev. E}, 74:031116, 2006.

\bibitem{GROMACS}
S.~Pronk, S.~P{\'a}ll, R.~Schulz, P.~Larsson, P.~Bjelkmar, R.~Apostolov, M.~R.
  Shirts, J.~C. Smith, P.~M. Kasson, D.~van~der Spoel, B.~Hess, and E.~Lindahl.
\newblock {GROMACS 4.5:} a high-throughput and highly parallel open source
  molecular simulation toolkit.
\newblock {\em Bioinformatics}, 29(7):845--854, 2013.

\bibitem{FaEl}
D.~Fange and J.~Elf.
\newblock Noise induced {M}in phenotypes in \textit{{E}.~coli}.
\newblock {\em PLoS Comput. Biol.}, 2(6):e80, 2006.

\bibitem{Kruse:2002}
K.~Kruse.
\newblock A dynamic model for determining the middle of \emph{Escherichia
  coli}.
\newblock {\em Biophys.~J.}, 82:618--627, 2002.

\end{thebibliography}
\providecommand{\noopsort}[1]{} \providecommand{\doi}[1]{\texttt{doi:#1}}
  \providecommand{\available}[1]{Available at \texttt{#1}}
  \providecommand{\availablet}[2]{Available at \texttt{#2}}

\end{document}